\documentclass[eprint,amsmath,amssymb,aps,nofootinbib]{revtex4-2}
\usepackage{mathtools,bm}
\usepackage{easyReview}
\usepackage{graphicx} 
\usepackage{xcolor}
\usepackage{hyperref}
\hypersetup{colorlinks=True,
  citecolor=olive,
  linkcolor=blue,
  urlcolor=black}  

\newcommand{\Xb}{\bm{X}}

\newcommand{\rb}{\bm{r}}
\newcommand{\fb}{\bm{f}}
\newcommand{\ub}{\bm{u}}

\newcommand{\xb}{\hat{\bm{x}}}
\newcommand{\yb}{\hat{\bm{y}}}
\newcommand{\eb}{\hat{\bm{e}}}
\newcommand{\tb}{\hat{\bm{t}}}
\newcommand{\nb}{\hat{\bm{n}}}
\newcommand{\zb}{\hat{\bm{z}}}

\newcommand{\consDiscreteHydro}{\widetilde{\bm{\mathrm{N}}}}

\newcommand{\discreteSlip}{\tilde{\bm{\mathrm{u}}}}
\newcommand{\bnabla}{\boldsymbol{\nabla}}
\newcommand{\disCenterline}{\bm{\mathrm{X}}}
\newcommand{\disTangent}{\bm{\tau}}

\begin{document}

\preprint{APS/123-QED}

\title{Instability and self-propulsion of flexible autophoretic filaments}

\author{Ursy Makanga\textcolor{blue}{$^\dagger$}}
\email{Contact author: umakanga@flatironinstitute.org}
\thanks{Present address: Center for Computational Biology, Flatiron Institute, Simons Foundation, New York, New York 10010, USA.}
\affiliation{Max Planck Institute for the Physics of Complex Systems, N\"othnitzer Stra{\ss}e 38,
01187 Dresden, Germany}

\author{Akhil Varma}
\thanks{These authors contributed equally to this work.}
\affiliation{Max Planck Institute for the Physics of Complex Systems, N\"othnitzer Stra{\ss}e 38,
01187 Dresden, Germany}

\author{Panayiota Katsamba}
\affiliation{Department of Chemical Engineering, Cyprus University of Technology, 30 Archbishop Kyprianou Str., Limassol 3036, Cyprus}
\affiliation{Computational-based Science and Technology Research Center (CaSToRC), The Cyprus Institute, Nicosia, 2121, Cyprus}

\date{\today}

\begin{abstract}
Over the past decade, autophoretic colloids have emerged as a prototypical system for studying self-propelled motion at microscopic scales, with promising applications in microfluidics, micromachinery, and therapeutics. Their motion in a viscous fluid hinges on their ability to induce surface slip flows that are spatially asymmetric from self-generated solute gradients. Here, we demonstrate theoretically that a straight elastic filament with homogeneous surface chemical properties -- which is otherwise immotile -- can spontaneously achieve self-propulsion by experiencing a buckling instability that serves as the symmetry-breaking mechanism. Using efficient numerical simulations, we characterize the nonlinear dynamics of the elastic filament and show that, over time, it attains distinct swimming modes such as a steadily translating “U” shape and a metastable rotating “S” shape when semiflexible, and an oscillatory state when highly flexible. Our findings provide physical insight into future experiments and the design of reconfigurable synthetic active colloids.
\end{abstract}

\maketitle

\section{Introduction}
\label{sec:introduction}
Ever since Antonie van Leeuwenhoek first peered through his handcrafted microscope in 1674 and observed “an abundance of very little and odd animalcules”, biological microswimmers such as bacteria, spermatozoa, and protozoa have captivated scientists with their ability to harness internal or environmental energy for self-propulsion in complex fluids. Nearly three centuries later, Richard Feynman promoted the idea of manipulating matter at the nanoscale to engineer functional micromachines \citep{feynman1991there}. This vision has partly guided scientists and engineers to devise ingenious ways of microscopic manipulation in the form of engineered, often bio-inspired, microswimmers and active colloids \citep{katuriArtificialMicroswimmersSimulated2016,guixSelfPropelledMicroNanoparticle2018,nsamelaColloidalActiveMatter2023}. Functionally, the latter have immense potential in medicine for targeted drug delivery and therapeutics \citep{nelsonMicrorobotsMinimallyInvasive2010,alapanMicroroboticsMicroorganismsBiohybrid2019,wuMedicalMicroNanorobots2020}, as well as in engineering in the form of micromachines for extracting work \citep{maggiSelfAssemblyMicromachiningSystems2016,aubretTargetedAssemblyMicrogears2018}, autonomous navigation \citep{simmchenTopographicalPathwaysGuide2016}, and cargo transport \citep{dreyfusMicroscopicArtificialSwimmers2005,sundararajanCatalyticMotorsTransport2008,barabanTransportCargoJanus2012}. More fundamentally, they are model active matter systems for exploring emergent behavior and collective dynamics \citep{zottlEmergentBehaviour2016,illienFuelledMotion2017}.

One such canonical system employs chemically active colloids that self-propel using a mechanism called \emph{autophoresis}, which is as follows: A catalytic coating on the particle's surface catalyzes a reaction that produces or depletes a solute in the surrounding fluid. Any inhomogeneities in this surface activity then create solute concentration gradients along its surface, which effectively produce surface slip flows through diffusiophoresis \citep{andersonColloidTransportInterfacial1989,derjaguinKineticPhenomenaBoundary1993}. Conservation of momentum then leads to the propulsion of the particle through the fluid. In this manner, it is clear that a symmetry breaking in the surface concentration field is necessary for an isotropically shaped particle to self-propel. Many mechanisms exist in the literature that provide the means to achieve this breaking. A classic one is a Janus colloid, an example of which is a silica (or gold) particle partially coated in platinum that catalyzes the breakdown of hydrogen peroxide in its solution \citep{paxtonCatalyticNanomotorsAutonomous2004,howseSelfMotileColloidalParticles2007}. This partial coating provides the necessary symmetry breaking of the surface concentration field \citep{golestanianPropulsionMolecularMachine2005, golestanianDesigningPhoreticMicro2007,moranPhoreticSelfPropulsion2017}. Another popular mechanism involves an advective instability of particles with homogeneous chemical surface properties \citep{michelinSpontaneousAutophoreticMotion2013,morozovNonlinearDynamicsChemicallyactive2019a}. It is observed in systems where the solute is large enough to be advected by the fluid, in a particular direction, faster than diffusion can homogenize it -- thereby breaking the symmetry. Chemically active droplets rely on surface concentration gradients of surfactant molecules for self-propulsion and typically function using a similar mechanism (Refs.\citep{maassSwimmingDroplets2016,michelinReviewActiveDroplets2023} and references therein). Yet another mechanism involves collective interactions of multiple particles that lead to the formation of geometrically asymmetric, self-propelling clusters \citep{varmaClusteringinducedSelfpropulsionIsotropic2018}.  

The shape anisotropy also significantly affects the characteristics of self-propulsion in active colloids \citep{shklyaevNonsphericalMotor2014,michelinAutophoreticGeometricAsymmetry2015}. The shapes analyzed initially were spheroids, cylinders, or disks \citep{golestanianDesigningPhoreticMicro2007,popescuPhoreticSpheroidal2010,nourhaniSelfelectrophoresisSpheroidalElectrocatalytic2015,michelinGeometricTuningSelfpropulsion2017}, with later designs having bent \citep{kummelCircularMotionAsymmetric2013,gangulyGoingCirclesSlender2023} and tori morphologies \citep{schmiedingAutophoreticTorus2017,bakerShapeprogrammed3DPrinted2019}. The slender morphology of rigid particles affords extra degrees of freedom and access to various swimming modes, such as stationary pumping (straight), translation (“U"-shaped), and rotation (“L" and “S"-shaped) \citep{kummelCircularMotionAsymmetric2013,sharanFundamentalModesSwimming2021,riedelDesigningHighlyEfficient2024, delmotteScalableMethodModel2024a}. First theoretical models of slender phoretic particles focused on straight filaments with spheroidal or arbitrary cross-sectional radius profiles \citep{solomentsevElectrophoresisSlenderParticles1994,yarivSlenderbodyApproximationsElectrophoresis2008,schnitzerOsmoticSelfpropulsionSlender2015,yarivSelfDiffusiophoresisSlenderCatalytic2020,saintillanHydrodynamicInteractionsInducedcharge2006}. Later, Katsamba \textit{et al.} \citep{katsambaSlenderPhoreticTheory2020} developed a \emph{slender phoretic theory} (SPT) -- that reduces the complexity of the phoretic problem from three dimensions to one -- using a matched asymptotic expansion of the boundary integral solution to Laplace's equation. The SPT framework tackles arbitrary three-dimensional geometries -- including open and closed loops -- with different cross-sectional radius and activity profiles \citep{katsambaChemicallyActiveFilaments2022,katsambaSlenderPhoreticLoops2024, KatsambaMontenegroJohnson2024_BookChapter}. Independently, Poehnl and Uspal \citep{poehnlPhoreticSelfpropulsionHelical2021} proposed a complementary source–dipole approach.      

While a majority of studies have focused on rigid active particles, flexible ones are recently gaining attention for their potential as multifunctional soft materials. Surface deformations typically require the material to have elasticity, which allows for shape changes in response to internal or external cues. Even in the absence of an internal actuation, the interplay of viscous and elastic tractions suffices to create rich nonlinear dynamics in passive flexible filaments. Examples include their planar bending and buckling \citep{guglielminiBucklingTransitionsElastic2012,liSedimentationFlexibleFilaments2013,evansElastocapillarySelffoldingBuckling2013}, stretch-coil transition \citep{youngStretchCoilTransitionTransport2007,kantslerFluctuations2012}, helicoidal buckling \citep{chelakkotFlowInducedHelicalCoiling2012,chakrabartiFlexibleFilamentsBuckle2020}, and other morphological transitions in background flow (Refs.\citep{dupratFluidStructureLowReynolds2015,duRoureDynamicsFlexibleFilaments2019} and references therein). Studies on the dynamics of passive filaments are often motivated by biological counterparts -- such as actin microfilaments and microtubules -- which are made “active" by the presence of ATP-driven molecular motors which impart internal stresses. Active filaments exhibit self-propulsion, bistable transitions, and spontaneous beating patterns  which serve important biological functions \citep{bourdieuActinFilamentVelocity1995,albertsEssentialCellBiology2014,vilfanFlagellaBeating2019,shiSustainedUnidirectionalRotation2022}. These instabilities are generally triggered by active compressive stresses and long-range hydrodynamic interactions, with thermal noise regulating the shape fluctuations \citep{jayaramanAutonomousMotilityActive2012,laskarHydrodynamicInstabilitiesProvide2013,chelakkotFlagellarBrownianParticles2014,laskarBrownianMicrohydrodynamicsActive2015,martin-gomezActiveBrownianFilaments2019}. Flexible active filaments have been modeled to varying degrees of detail and complexity; some of which include bead chains driven by nonreciprocal interactions \citep{jayaramanAutonomousMotilityActive2012,laskarHydrodynamicInstabilitiesProvide2013,laskarBrownianMicrohydrodynamicsActive2015,mannaColloidalTransportActive2017,nemethNonReciprocalActiveSolids2025,al-izziNonreciprocalBucklingMakes2026a} and slender bodies with a follower force \citep{decanioSpontaneousOscillationsElastic2017,lingInstabilityElasticFilament2018, manMorphologicalTransitionsAxiallydriven2019,steinSwirlingInstabilityMicrotubule2021,clarkeBifurcationsNonlinearDynamics2024,schnitzerOnsetSpontaneousBeating2025}, with patterned active forces and moments \citep{youngFlexiblePolarFilament2010,loughSelfbucklingSelfwrithingSemiflexible2023}, or with more complex coupled-interactions \citep{camaletGenericAxonemalBeating2000,chakrabartiSpontaneousOscillations2019}.

Early prototypes of synthetic, flexible active filaments include magnetic filaments and colloidal chains which are actuated by an external electric or magnetic field \citep{dreyfusMicroscopicArtificialSwimmers2005,nishiguchiFlagellarJanusElectric2018,huangAdaptiveLocomotionArtificial2019,yangReconfigurableMicrobotColloidal2020, biswasEmergentSofteningStiffening2025, weiLifelikeBehaviorEmerging2026}. They exhibit nonreciprocal beating patterns and rotational dynamics, reminiscent of ciliary or flagellar motion. They have inspired phoretic analogs -- which are self-actuated -- such as autocatalytic colloidal bead chains \citep{biswasLinkingIsotropicColloids2017,vutukuriRationalDesignDynamics2017} and polymeric active droplets \citep{kumarEmergentDynamicsDue2023}.  Introducing self-actuation via chemical activity in flexible filaments fundamentally alters their dynamical landscape, giving rise to spontaneous symmetry breaking and emergent motility. Furthermore, models of chemoresponsive materials have been shown to enable tunable shape morphing through concentration-dependent mechanics which yield pumping, self-propulsion, and oscillations \citep{montenegro-johnsonMicrotransformersControlledMicroscale2018,mannaHarnessingPowerActiveSheets2022,qiaoControlActivePolymeric2022,altunkeyikActivePoroelasticFilaments2025}. In all these cases, flexibility enables dynamic shape modulation which allows switching between the swimming modes. This important additional feature presents a compelling incentive for detailed modeling of these synthetic active systems. 

From the numerous works discussed above, it is evident that active stresses induce rich nonlinear dynamics in flexible filaments, with significant practical implications. Motivated by these findings, we explore a scenario in which autophoretic effects are the source of internal active stresses. Among the many possible surface patterns that result in complex chemoelastohydrodynamic coupling, here we investigate the simple yet insightful case of filaments with homogeneous chemical surface properties. We describe a new mechanism in which a flexible, chemically isotropic filament that is otherwise stationary achieves self-propulsion in a viscous fluid due to a buckling instability. 
Recently, Butler \textit{et al.} \citep{butlerElastohydrodynamicsThreedimensionalChemically2026}\footnote{We note that we have been in communication with the team of Matthew Butler, Benjamin Walker, Tom Montenegro-Johnson, and Panayiota Katsamba (also an author of the present manuscript), who have recently submitted a complementary work titled “Elastohydrodynamics of three-dimensional chemically-active filaments” to the \textit{Journal of Fluid Mechanics}. While the present work was developed independently, we ensured that there is no significant overlap with their study after we became aware of it.} integrated SPT with an elastohydrodynamic solver \citep{walkerEfficientSimulationFilament2020a} to simulate the three-dimensional chemoelastohydrodynamics of self-actuated filaments with nonhomogeneous surface chemical properties. Their numerical study shows that the filaments undergo a buckling instability and reports a rich array of dynamic behaviors ranging from planar when semiflexible to out-of-plane conformations when highly flexible. For readers interested in a parametric study, we redirect them to the results from their simulations. In contrast, here we investigate theoretically the mechanism of the instability and characterize the resulting planar swimming modes via numerical simulations and scaling analysis -- aspects that are not covered in Ref.\citep{butlerElastohydrodynamicsThreedimensionalChemically2026}. 

The paper is organized as follows. We describe the geometry of the filament in Section \ref{section:filamentGeometry}. The dynamical equations for the chemoelastohydrodynamics of the filament are systematically derived using slender body approximations in Section \ref{section: chemo_elasto_hydro_problem}. We then reformulate the governing equations into a tangent-based form that later allows us to implement an efficient numerical approach via a pseudospectral method in Section \ref{section: numerical_methods} for solving the governing equations. Using the case of a uniformly active filament, we perform a linear stability analysis in Section \ref{section: linear_stability_analysis} to show that the filament experiences a buckling instability beyond a critical active stress. The buckled configurations provide the symmetry breaking necessary for self-propulsion. After validating our numerical approach near the instability threshold, simulations are carried out in Section \ref{section:nonlinearDynamics} to study the nonlinear dynamics of the flexible filament, thereby identifying and characterizing the various modes of self-propulsion. Conclusions are drawn in Section \ref{section:conclusions}.

\section{Filament geometry}
\label{section:filamentGeometry}
\begin{figure}[ht]
    \centering
    \includegraphics[width=0.73\linewidth]{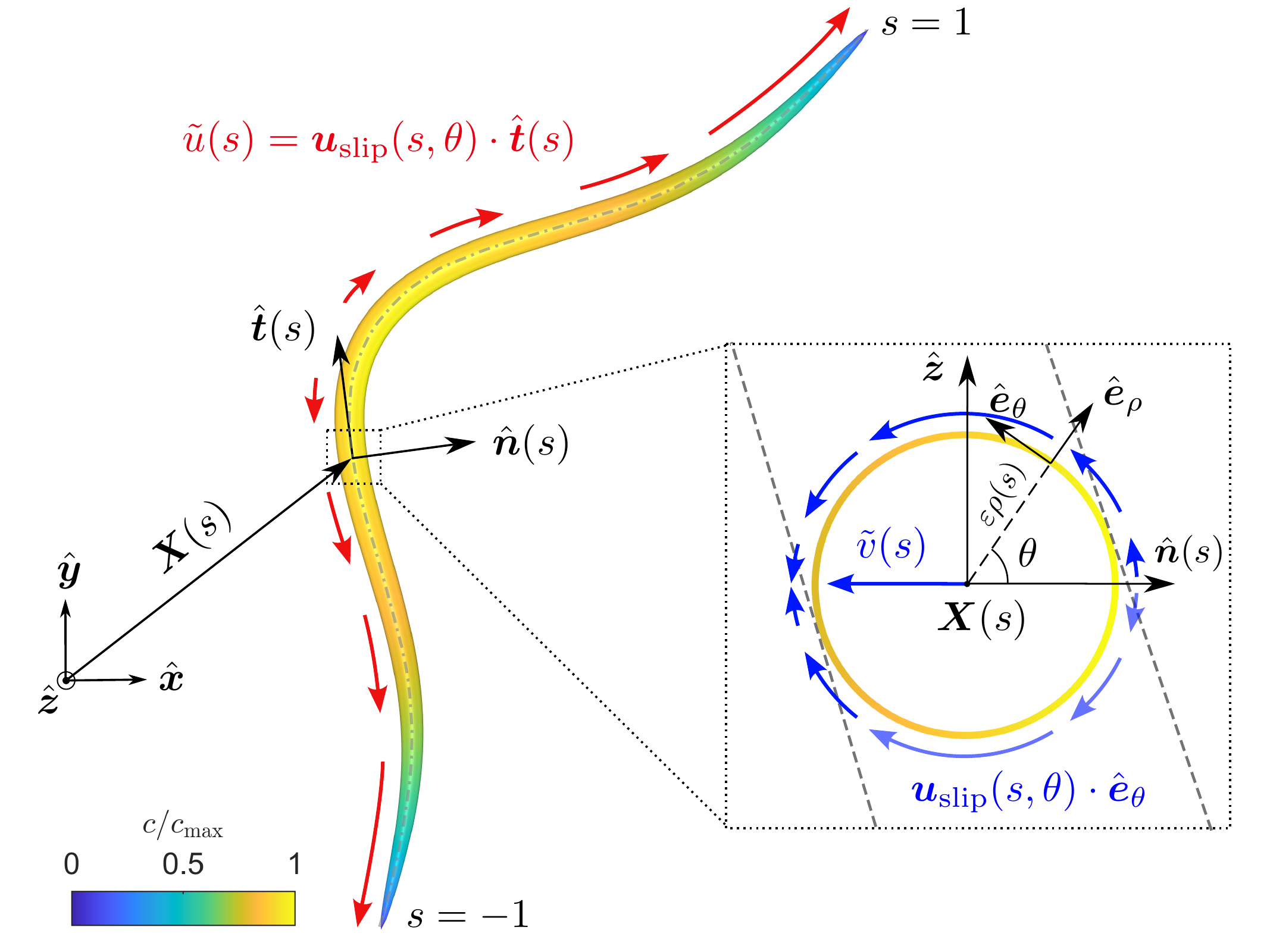}
    \caption{Schematic of the planar deformation of an autophoretic filament with uniform surface activity $\mathcal{A}$ and phoretic mobility $\mathcal{M}$. The centerline of the filament is defined by the position vector $\Xb(s)$ which is parameterized by the arc length coordinate $s$. The tangent and normal unit vectors, respectively $\tb(s)$ and $\nb(s)$, are attached to the centerline to keep track of the deformation. The surface concentration $c(s, \theta)$ and the resulting slip velocity $\bm{u}_\mathrm{slip}(s, \theta)$ are shown for $\varepsilon = 10^{-2}$ and a prolate spheroidal profile, $\rho(s)=\sqrt{1-s^2}$, having a positive activity (solute emission) $\mathcal{A}=1$ and a negative phoretic mobility $\mathcal{M}=-1$. The slip velocity is a superposition of the tangential $\Tilde{u}\tb=(\bm{u}_\mathrm{slip} \cdot \tb)\tb$ and azimuthal  $(\bm{u}_\mathrm{slip} \cdot \hat{\bm{e}}_\theta)\hat{\bm{e}}_\theta$ slip velocities. The latter is effectively described as a normal vector by its circumferential average, $\tilde{v}\nb$ (inset).}
    \label{fig:schematic}
\end{figure}

We begin by describing the geometry of the filament, illustrated in Fig.\ref{fig:schematic}. The position of the centerline of the filament at any given time $t$ is denoted by $\bm{X}(s,t)$. It is parameterized by the arc length coordinate, i.e., Lagrangian parameter,  $s \in [-l, l]$, where $2l$ is the total contour length.

In this study, we assume planar deformations (in the $xy$ plane) of the filament, i.e., twisting of the centerline does not occur. Thus, to keep track of the deformation, we introduce the orthonormal frame $(\hat{\bm{t}}(s,t), \hat{\bm{n}}(s,t))$, attached to the centerline of the filament, such that the tangent $\hat{\bm{t}}(s,t)$ and normal  $\hat{\bm{n}}(s,t)$ unit vectors satisfy the Serret-Frenet formulas,

\begin{equation}
    \frac{\partial \hat{\bm{t}}}{\partial s} = \kappa  \hat{\bm{n}} \quad \mathrm{and} \quad 
    \frac{\partial \hat{\bm{n}}}{\partial s} = -\kappa \hat{\bm{t}},
    \label{eq: serret_frenet_formulas}
\end{equation}
where $\hat{\bm{t}} =\partial_s \Xb$ and $\kappa(s, t)$ is the local curvature of the centerline. In the following, we omit writing the time dependence explicitly for notation convenience.

Assuming a circular cross section, we denote by $\bm{S}(s,\theta)$ the position of a point on the surface of the filament parameterized by $s$ and the azimuthal angle $\theta \in [-\pi, \pi]$,

\begin{equation}
    \bm{S}(s,\theta) = \bm{X}(s) + a \rho(s) \hat{\bm{e}}_\rho(s, \theta),
    \label{eq: position_of_points_on_surface}
\end{equation}
where $a \rho(s)$ is the cross-sectional radius profile, with $a$ being the maximal radius and $\rho(s) \in [0, 1]$. The radially outward-pointing unit vector, $\hat{\bm{e}}_\rho$, and azimuthal unit vector, $\hat{\bm{e}}_\theta$, form a local basis that is attached to the surface of the filament at any cross section,

\begin{equation}
    \hat{\bm{e}}_\rho = \cos{\theta}\hat{\bm{n}} + \sin{\theta}\hat{\bm{z}}\quad \mathrm{and} \quad
    \hat{\bm{e}}_\theta = -\sin{\theta}\hat{\bm{n}} + \cos{\theta}\hat{\bm{z}},
    \label{eq: local_basis_cross_section}
\end{equation}
where the unit vector $\zb$ is normal to the filament plane and is defined as $\zb=\xb\times\yb$. In what follows, we scale lengths upon $l$ and introduce the slenderness parameter, i.e., aspect ratio, $\varepsilon = a/l$. The arc length coordinate is then $s \in [-1,1]$, and the cross-sectional radius profile is $\varepsilon\rho(s)$. Figure \ref{fig:schematic} shows a prolate spheroidal profile, $\rho(s)=\sqrt{1-s^2}$.

\section{Dynamical equations of flexible autophoretic filaments}
\label{section: chemo_elasto_hydro_problem}
In this section, we derive the governing equations of the dynamics of flexible autophoretic filaments in a viscous fluid. The filament catalyzes a chemical reaction on its surface, emitting or absorbing a certain solute at a rate $\bar{\mathcal{A}}$, known as \emph{activity}; a positive (respectively, negative) value of activity indicates solute emission (respectively, absorption). Furthermore, the solute molecules interact with the filament surface, inducing flows due to concentration gradients within a thin boundary layer close to the surface -- a mechanism known as diffusiophoresis \citep{andersonColloidTransportInterfacial1989, derjaguinKineticPhenomenaBoundary1993}. At the scale of the filament, the flow is generally considered as an effective slip on its surface, having a velocity that is proportional to the local surface gradients of the concentration; the coefficient of proportionality is a physicochemical property of the surface known as \emph{mobility}, $\bar{\mathcal{M}}$. A surface which possesses both the properties of activity and mobility, as in the present case, is termed \emph{autophoretic} \citep{golestanianPropulsionMolecularMachine2005}.

\subsection{Autophoretic problem}
Let $\bar{c}(\bm{r})$ be the concentration of the solute at any point $\rb$ in the fluid domain, relative to the equilibrium concentration far from the filament. The characteristic surface concentration of the filament is $|A|a/D$, where $A$ and $D$ are the characteristic activity of the filament and the diffusivity of the solute, respectively. We assume that the solute diffuses rapidly, much faster than advection. Then the solute concentration, after nondimensionalizing using the characteristic value, obeys Laplace's equation
\begin{equation}
    \nabla^2 c(\bm{r}) = 0,
    \label{eq: laplace_equation}
\end{equation}
where $c$ is the dimensionless solute concentration.

In dimensionless terms, the solute flux on the surface of the filament is
\begin{align}
   \bm{n}_f \cdot \bnabla c(\bm{r}) = -\mathcal{A}(\bm{r}) \quad \mathrm{for} \quad \bm{r} = \bm{S}(s, \theta),
   \label{eq: solute_flux_surface_filament}
\end{align}
where $\bm{n}_f$ is an outward-pointing unit vector normal to the surface of the filament and $\mathcal{A}(\bm{S})=\bar{\mathcal{A}}(\bm{S})/|A|$ is the dimensionless activity. Far from the filament, the solute concentration field vanishes, 

\begin{equation}
    c(\bm{r}) = 0 \quad \mathrm{as} \quad |\bm{r}| \rightarrow \infty.
    \label{eq: solute_concentration_infinity}
\end{equation}

In what follows, unless otherwise stated, we consider filaments with axisymmetric activity, $\bar{\mathcal{A}}(\bm{S}) \equiv \bar{\mathcal{A}}(s)$, and phoretic mobility, $\bar{\mathcal{M}}(\bm{S}) \equiv \bar{\mathcal{M}}(s)$. For a slender filament, $\varepsilon \ll 1$, the solution to the phoretic problem, outlined by Eqs.\eqref{eq: laplace_equation}-\eqref{eq: solute_concentration_infinity}, was derived by Katsamba \textit{et al.} \citep{katsambaSlenderPhoreticTheory2020} and referred to as the \emph{slender phoretic theory}  (SPT). This approach asymptotically reduces the phoretic problem from three dimensions to one, resulting in a distribution of source monopoles and dipoles along the centerline of the filament. The idea of SPT is to assume an asymptotic expansion of the surface concentration in powers of the slenderness parameter, $\varepsilon$, of the form
\begin{equation}
    c(s,\theta)=c^{(0)} (s) +\varepsilon c^{(1)}(s,\theta) + \mathcal{O}(\varepsilon^2),
\end{equation}
where, by retaining only the local terms, the leading-order surface concentration, $c^{(0)}$, is given by
\begin{align}
    c^{(0)}(s) &=  \frac{1}{2}b(s)\rho(s)\mathcal{A}(s)
    \label{eq:surf_concentration_c0}
\end{align}
and the first-order surface concentration,  $c^{(1)}$, by
\begin{align}
c^{(1)}(s, \theta) = \frac{1}{2}\left[b(s)-3\right]\rho(s)^2\kappa(s)\mathcal{A}(s)\cos{\theta} 
    \label{eq:surf_concentration_c1}
\end{align}
with $b(s) = \log (4(1-s^2)/(\varepsilon^2 \rho(s)^2))$. For a given filament activity, radius, and curvature, these local terms give the surface concentration contribution from points $\mathcal{O}(\varepsilon)$ away from $\bm{X}(s)$. Note that an improper choice of the cross-sectional radius profile $\rho(s)$ or the activity $\mathcal{A}(s)$ can result in singularities at the endpoints, $s=\pm 1$. A prolate spheroidal profile $\rho(s)= \sqrt{1 - s^2}$ or a smooth activity profile $\mathcal{A}(s)=s^2-1$ can be used to circumvent this problem.  

The gradients in concentration will give rise to a phoretic slip velocity on the surface of the filament, $\bm{u}_\mathrm{slip}(\bm{r}) = \bar{\mathcal{M}}(s) \bnabla_{\parallel} \bar{c}(\bm{r})$ for $\bm{r}=\bm{S}(s, \theta)$, where $\bnabla_{\parallel}[\bullet] = \partial_s [\bullet]\tb + 1/(a\rho(s)) \partial_\theta [\bullet]\eb_{\theta}$
denotes the surface gradient \citep{katsambaSlenderPhoreticTheory2020}. The characteristic slip velocity is then $V=|AM|a/(Dl)$, where $M$ is the characteristic phoretic mobility of the filament. Note that this is only a fraction ($a/l = \varepsilon \ll 1$) of the characteristic slip velocity of a spherical colloid of the same diameter \citep{golestanianDesigningPhoreticMicro2007}. Since the swimming velocity is constant on the cross section up to $\mathcal{O}(\varepsilon)$ \citep{katsambaSlenderPhoreticTheory2020}, the circumferential average of the slip velocity, $\tilde{\bm{u}}_{\mathrm{slip}}(s)$, is therefore the leading-order contribution to the kinematics of the filament. Then, by averaging $\bm{u}_\mathrm{slip}$ over $\theta$ and after nondimensionalizing using the characteristic value, we obtain
\begin{equation}
    \tilde{\bm{u}}_{\mathrm{slip}}(s) = \tilde{u}(s) \tb(s) + \tilde{v}(s)\nb(s),
    \label{eq: phoretic_slip_velocity}
\end{equation}
with the components being
\begin{subequations}
    \label{eq: phoretic_slip_velocity_components}
    \begin{align}
    \tilde{u}(s) &= \mathcal{M}(s) \partial_s c^{(0)}(s) \quad \mathrm{and} 
    \label{eq:slip_tangent}\\
    \tilde{v}(s) &= \frac{1}{4}[b(s)-3] \rho(s) \kappa(s) \mathcal{A}(s)\mathcal{M}(s),
    \label{eq:slip_normal}
    \end{align}
\end{subequations}
where $\mathcal{M}(s)=\bar{\mathcal{M}}(s)/|M|$ is the dimensionless phoretic mobility. Thus $\tilde{\bm{u}}_{\mathrm{slip}}$ comprises two distinct contributions: (i) a tangential component, $\tilde{u}$, arising from gradients of $c^{(0)}$ along the filament length, independent of the filament configuration, and (ii) a normal component, $\tilde{v}$, resulting from gradients of $c^{(1)}$ across the filament cross section, proportional to the curvature.

\subsection{Filament elasticity}
We model the filament as a linearly elastic material, i.e., its elastic response remains linear under a given stress. For a slender filament, $\varepsilon \ll 1$, that is immersed in a viscous fluid, the timescales at which it twists and stretches are negligible compared to the timescale at which it bends, for a moderate forcing \citep{audolyElasticityGeometryHair2010}. Therefore, we consider only planar deformations of the filament, neglecting twisting and stretching. Under these assumptions, the elastic force density is given by Euler-Bernoulli beam theory for thin elastic rods \citep{landauTheoryElasticityVolume1986, audolyElasticityGeometryHair2010}
\begin{equation}
    \fb_\mathrm{elastic}(s) = \partial_s(T(s)\partial_s \bm{X}(s)) - \partial_s^2(B(s)\partial_s^2\bm{X}(s)),
    \label{eq:elastic_force_density}
\end{equation}
where $T$ is the internal line tension which acts as a Lagrange multiplier that enforces the inextensibility condition,

\begin{equation}
    \partial_s \bm{X}(s) \cdot \partial_s \bm{X}(s) = 1
    \label{eq: inextensibility_condition}.
\end{equation}
Here $B(s)=EI(s)$ is the bending rigidity, with $E$ and $I(s)=\pi a^4 \rho(s)^4/4$ being the Young's modulus and the second moment of inertia, respectively. In this study,  we assume a uniform bending rigidity, i.e., $\partial_sB(s)=0$; and the ends of the filament are taken to be free, i.e., no external forces and torques are applied on them. Therefore, the following boundary conditions are satisfied at both ends of the filament \citep{liSedimentationFlexibleFilaments2013} 

\begin{align}
   \partial_s^2\bm{X}(s)|_{s=\pm 1} = \bm{0},\quad \partial_s^3\bm{X}(s)|_{s=\pm 1} = \bm{0}, \;\; \mathrm{and}\quad T(s)|_{s=\pm 1}=0.
    \label{eq: boundary_conditions}
\end{align}

\subsection{Chemoelastohydrodynamic problem} 
\label{sec:Chemo_elastohydrodynamic_problem}
The phoretic filament being immersed in an incompressible viscous fluid of viscosity $\eta$, in the absence of inertia, the fluid flow obeys the Stokes equations

\begin{equation}
    -\eta\nabla^2 \ub + \bnabla p = \bm{0}, \qquad \bnabla \cdot \ub = 0,
\end{equation}
where $\bm{u}$ and $p$ are the fluid velocity and pressure, respectively. Assuming that the filament is moving in an infinite quiescent fluid domain, the boundary condition far from the surface of the filament is $\bm{u}(\bm{r})= \bm{0}$ as $|\bm{r}| \rightarrow \infty$. On the surface of the filament, the fluid velocity is a superposition of the swimming and phoretic slip velocities, $\bm{u}(\bm{r}) = \partial_t \bm{r} + \bm{u}_{\mathrm{slip}}(\bm{r})$  for $\bm{r} = \bm{S}(s, \theta)$. For a slender filament, $\varepsilon \ll 1$, the fluid velocity, on the surface, is given by the \emph{slender body theory} (SBT). SBT is an asymptotic approach that relies on a superposition of singularities to construct solutions for the flow around a moving slender object. This approach was first introduced by Hancock \citep{hancockSelfpropulsionMicroscopicOrganisms1953} and later extended at different degrees of approximation \citep{coxMotionLongSlender1970, batchelor_slender-body_1970, lighthillFlagellarHydrodynamics1976a, kellerSlenderbodyTheorySlow1976,johnsonImprovedSlenderbodyTheory1980, gotzDissertation2001, cortezSlenderBodyTheory2012, koensBoundaryIntegralFormulation2018, ohmMathematicalFoundationsSlender2020a, maxianIntegralbasedSpectralMethod2021}. In this study, we use the SBT approach of Johnson \citep{johnsonImprovedSlenderbodyTheory1980} that approximated the fluid velocity on the cross section of the filament up to $\mathcal{O}(\varepsilon)$.    

Scaling lengths upon $l$, forces upon $B/l^2$, velocities upon $V$, times upon the diffusiophoretic timescale $\tau_\mathrm{phoretic} = l/V$, and introducing the timescale for elastic relaxation $\tau_\mathrm{elastic} = 8\pi\eta l^4/B$; the circumferential average of the fluid velocity, on the surface of the filament, reads

\begin{equation}
    \alpha \left[\partial_t{\Xb}(s)+ \tilde{\ub}_{\mathrm{slip}}(s)\right] = \bm{\mathcal{N}}[\fb_\mathrm{elastic}](s),
    \label{eq:SBT1}
\end{equation}
where the dimensionless elastic force density is $\fb_\mathrm{elastic}(s) = \partial_s(T(s)\partial_s \bm{X}(s)) - \partial_s^4\bm{X}(s))$. By retaining only the local term, the SBT hydrodynamic operator $\bm{\mathcal{N}}$ is defined such that

\begin{align}
    \bm{\mathcal{N}}[\fb_\mathrm{elastic}](s) =  b(s) [\bm{I} + \tb(s) \tb(s)]\cdot \fb_\mathrm{elastic}(s) + [\bm{I}-3\tb(s)\tb(s)] \cdot \fb_\mathrm{elastic}(s),
\end{align}
with $\bm{I}$ being the identity tensor. 

We see in Eq.\eqref{eq:SBT1} that the filament dynamics is described by a single dimensionless parameter,
\begin{equation}
    \alpha = \frac{8\pi\eta a l^2|AM|}{B D} \equiv \frac{\tau_\mathrm{elastic}}{\tau_\mathrm{phoretic}},
    \label{eq:alpha}
\end{equation}
termed as the \emph{elastophoretic number}. The latter compares the timescale for elastic relaxation with the timescale for diffusiophoretic flow and is equivalent to the elastoviscous number: $\alpha \equiv 8\pi \eta V l^3/B$. For notation convenience, we will write Eq.\eqref{eq:SBT1} abstractly as

\begin{equation}
    \alpha \left[\partial_t {\Xb}(s) + \tilde{\ub}_{\mathrm{slip}}(s)\right]\vcentcolon = \bm{\mathcal{N}}[\bm{X}]\cdot \left[ \partial_s(T(s) \partial_s\Xb(s)) - \partial_s^4\Xb(s)\right],
    \label{eq:SBT_final}
\end{equation}
for the remainder of this study. Here, the hydrodynamic mobility operator $\bm{\mathcal{N}}[\bm{X}]$ is a functional of the centerline position, $\Xb$.

\subsection{Summary of the dynamical equations}
We summarize here the equations governing the dynamics of a flexible autophoretic filament in a viscous fluid. These equations are given by the kinematic equation of motion \eqref{eq:SBT_final},

\begin{equation}
    \partial_t \bm{X}(s) = \bm{\mathcal{N}}[\Xb]\cdot \left[ \partial_s(T(s)\partial_s \bm{X}(s)) -\frac{1}{\alpha} \partial_s^4\bm{X}(s) \right] - \tilde{\ub}_{\mathrm{slip}}(s),
    \label{eq:SBT_final_with_tension}
\end{equation}
together with the inextensibility condition \eqref{eq: inextensibility_condition},

\begin{equation}
    \partial_s \bm{X}(s) \cdot \partial_s \bm{X}(s) = 1
    \label{eq: inextensibility_condition_summarized},
\end{equation}
and the boundary conditions \eqref{eq: boundary_conditions},

\begin{align}
   \partial_s^2\bm{X}(s)|_{s=\pm 1} = \bm{0},\quad \partial_s^3\bm{X}(s)|_{s=\pm 1} = \bm{0}, \;\; \mathrm{and}\quad T(s)|_{s=\pm 1}=0.
    \label{eq: boundary_conditions_summarized}
\end{align}
Note that since the internal line tension, $T$, is unknown, it has been rescaled by $\alpha$ in Eq.\eqref{eq:SBT_final_with_tension} without any loss of generality. 

In the present model, the chemical species were assumed to diffuse much faster than both advection by the flow (P\'eclet number, $\mathrm{Pe} \ll 1$) and shape changes of the filament, allowing the concentration field to be treated as quasisteady for each instantaneous filament configuration. Similarly, the flow field is considered to be quasisteady, as the effects of fluid inertia are negligible compared to viscous ones (Reynolds number, $\mathrm{Re} \ll 1$). Additionally, Eqs.\eqref{eq:SBT_final_with_tension}-\eqref{eq: boundary_conditions_summarized} were derived assuming planar deformations of the filament and uniform bending rigidity. Equation \eqref{eq:SBT_final_with_tension} follows from the slender body approximations ($\varepsilon \ll 1$) of both the hydrodynamic and chemical problems, namely, respectively, local SBT and SPT. The latter are both valid approximations as long as the filament is locally straight, i.e., $\kappa \ll \varepsilon^{-1}$. Nonlocal contributions can however become important when the filament exhibits large global deformations when extremely flexible ($\alpha \to \infty$), which is not presently accounted for.

The numerical approach to solve these dynamical equations can be determined by its effectiveness in the numerical treatment of the inextensibility condition. The traditional approach to handle the latter, known as the \textit{tension-based formulation} \citep{tornbergSimulatingDynamicsInteractions2004}, relies on an integrodifferential equation for the internal line tension. Despite the use of this formulation in several studies \citep{liSedimentationFlexibleFilaments2013, manikantanBucklingTransitionSemiflexible2015, nazockdastFastPlatformSimulating2017, chakrabartiFlexibleFilamentsBuckle2020}, it leads to some drawbacks such as the loss of spatial accuracy when using spectral methods \citep{nazockdastFastPlatformSimulating2017, maxianHydrodynamicsTransientlyCrossLinked2023}; furthermore some penalty terms are required to preserve inextensibility \citep{tornbergSimulatingDynamicsInteractions2004}. In this study, we use an alternative approach which has been used in recent studies (Refs.\citep{moreauAsymptoticCoarsegrainingFormulation2018,hall-mcnairEfficientImplementationElastohydrodynamics2019, walkerFilamentMechanicsHalfspace2019, jabbarzadehNumericalMethodInextensible2020,maxianIntegralbasedSpectralMethod2021}, to cite a few) to overcome these hurdles -- the \textit{tangent-based formulation}.

\subsection{Tangent-based formulation}
In the tangent-based formulation, the filament dynamics is evolved through the tangent vector $\hat{\bm{t}}$ rather than the centerline position $\bm{X}$. Accordingly, the time derivative of the inextensibility condition \eqref{eq: inextensibility_condition_summarized} implies that

\begin{equation}
\partial_t{\bm{\hat{t}}}(s) = \bm{\Omega}(s) \times \bm{\hat{t}}(s),
\label{eq: time_derivative_tangent_vector}
\end{equation}
where $\bm{\Omega}$ is an angular velocity. The filament evolution can be thought of as rotations of the tangent vector on unit sphere. 

From the relation $\partial_s \bm{X}(s) = \hat{\bm{t}}(s)$, the centerline position can be written as 

\begin{equation}
    \bm{X}(s) = \bm{X}(s)|_{s=0} + \int_{0}^{s} \hat{\bm{t}}(s') \mathrm{d}s',
    \label{eq: robot_arm}
\end{equation}
where $\bm{X}(s)|_{s=0}$ denotes the evaluation of $\bm{X}$ at the midpoint, $s=0$. Then, by taking the time derivative of Eq.\eqref{eq: robot_arm} and using Eq.\eqref{eq: time_derivative_tangent_vector}, we express the velocity of the centerline in terms of the angular velocity of the tangent vector, $\bm{\Omega}$, and the linear velocity at the midpoint, $\bm{U}(s)|_{s=0}$, 

\begin{equation}
  \partial_t \bm{X}(s) = \bm{U}(s)|_{s=0} + \int_{0}^s \left[\bm{\Omega}(s') \times \bm{\hat{t}}(s') \right]\mathrm{d}s' \vcentcolon = \bm{\mathcal{K}}[\bm{X}]\cdot \bm{\phi}(s), 
\end{equation}
where $\bm{\phi}(s) = \{ \bm{\Omega}(s), \bm{U}(s)|_{s=0}\}$, and the kinematic operator $\bm{\mathcal{K}}[\bm{X}]$ restricts the filament configuration $\bm{X}$ onto the space of kinematically admissible configurations. Accordingly, the kinematic equation of motion \eqref{eq:SBT_final_with_tension} can be rewritten as 
    
\begin{equation}
    \partial_t \bm{X}(s) = \bm{\mathcal{N}}[\Xb]\cdot \left[ \bm{\lambda}(s) + \bm{\mathcal{F}} \cdot \bm{X} \right] - \tilde{\ub}_{\mathrm{slip}}(s) \vcentcolon = \bm{\mathcal{K}}[\bm{X}]\cdot \bm{\phi}(s),
    \label{eq:SBT_final_with_lambda}
\end{equation}
where $\bm{\lambda}(s) = \partial_s(T(s)\partial_s \bm{X}(s))$ is the constraint force that enforces the inextensibility condition \eqref{eq: inextensibility_condition} and $\bm{\mathcal{F}}$ is a linear operator that gives the bending force density, $\bm{\mathcal{F}} \cdot \bm{X} \vcentcolon = -\delta\mathcal{E}^B/\delta \bm{X}$, where the bending energy functional $\mathcal{E}^B[\bm{X}]$ is given by

\begin{equation}
    \mathcal{E}^B[\bm{X}] = \frac{1}{2\alpha}\int_{-1}^{1} \left\lVert\partial_s^2\bm{X}(s)\right\rVert^2_2 \;\mathrm{d}s,
\end{equation}
with  $\lVert \bullet \rVert_2$ being the $L^2$ norm. This energy-based formulation of the bending force density is suitable for numerical simulations since it implicitly enforces the free-end boundary conditions,  $\partial_s^2\bm{X}(s)|_{s=\pm 1} = \bm{0}$ and $\partial_s^3\bm{X}(s)|_{s=\pm 1} = \bm{0}$.  

In the resulting kinematic equation of motion \eqref{eq:SBT_final_with_lambda}, the constraint force $\bm{\lambda}$ and the velocities $\bm{\phi}$ are both unknowns. Since the constraint force is workless, the tangent-based formulation can be closed by a kinematic constraint equation that is derived via the principle of virtual work \citep{maxianIntegralbasedSpectralMethod2021}, 

\begin{equation}
\bm{\mathcal{K}}^\dagger[\bm{X}]\cdot \bm{\lambda}(s) \vcentcolon = 
\begin{bmatrix}
\displaystyle\int_{-1}^{1} \bm{\lambda}(s)\;\mathrm{d}s\\\\
\bm{\hat{t}}(s) \times \displaystyle\int_{s}^{1} \bm{\lambda}(s')\;\mathrm{d}s' 
\end{bmatrix}
=
\begin{bmatrix}
\vphantom{\displaystyle\int_{-1}^{1} \bm{\lambda}(s)\;\mathrm{d}s
}
  \bm{0} \\\\
 \bm{0}
\vphantom{\displaystyle\int_{-1}^{1} \bm{\lambda}(s)\;\mathrm{d}s
}
\end{bmatrix},
\label{eq: kinematic_constraint_virtual_work}
\end{equation}
where $\bm{\mathcal{K}}^\dagger[\bm{X}]$ is the $L^2$ adjoint operator of $\bm{\mathcal{K}}[\bm{X}]$. The above equation gives the boundary conditions $T(s)|_{s=\pm 1}=0$ and, therefore, enforces the force-free condition, $\int_{-1}^{1} [\partial_s(T(s)\partial_s \bm{X}(s)) -\alpha^{-1}\partial_s^4\bm{X}(s)]=\bm{0}$. Its second row corresponds to the moment induced by the internal force $\int_{s}^{1} \bm{\lambda}(s')\;\mathrm{d}s'$ at $s$ and implies that $\bm{\lambda}(s) = \partial_s(T(s)\partial_s \bm{X}(s))$.

Thus, the filament motion is determined by the continuous solution $\{\bm{\lambda}, \bm{\phi}\}$  to the following saddle point system: 

\begin{equation}
    \begin{bmatrix}
    -\bm{\mathcal{N}} &\vphantom{\ldots} & \bm{\mathcal{K}} \\\\
    \bm{\mathcal{K}}^\dagger                               & &\vphantom{\ldots} \bm{0}\\
    \end{bmatrix}
    \cdot
    \begin{bmatrix}
    \bm{\lambda}\\\\
    \bm{\phi}
    \end{bmatrix}
    = 
    \begin{bmatrix}
    \bm{\mathcal{N}} \cdot \bm{\mathcal{F}} \cdot \bm{X} - \tilde{\ub}_{\mathrm{slip}}\\\\
    \bm{0}
    \end{bmatrix}
    \label{eq: mixed_mob_res_problem},
\end{equation}
which results from Eqs.\eqref{eq:SBT_final_with_lambda} and \eqref{eq: kinematic_constraint_virtual_work}.

\section{Numerical methods}
\label{section: numerical_methods}
We numerically solve the dynamical equations, through the tangent-based formulation, using a pseudospectral method. We summarize the method here, while a detailed description can be found elsewhere \citep{maxianIntegralbasedSpectralMethod2021, maxianHydrodynamicsTransientlyCrossLinked2023, maxianSimulationPlatformSlender2024b}.

\subsection{Spatial discretization}
We represent the tangent vector $\hat{\bm{t}}$ and the centerline position $\bm{X}$ in Chebyshev bases of the first kind with $N$ and $N_X = N+1$ (including the midpoint, $\bm{X}(s)|_{s=0}$) nodes, respectively. For notation convenience, we will use the same symbol for the true tangent vector $\hat{\bm{t}}$ and its corresponding interpolant, and likewise for the centerline position $\bm{X}$. We denote by $\disTangent = \{\hat{\bm{t}}(s_p)\}_{p=1}^{N} \in \mathbb{R}^{3N}$ and $\disCenterline = \{\bm{X}(s_p)\}_{p=1}^{N_X} \in \mathbb{R}^{3N_X}$ the vectors collecting the tangent vectors and centerline positions at the collocation points $s_p$, respectively; by  $\discreteSlip \in \mathbb{R}^{3N_X}$ the discrete description of the phoretic slip velocity $\tilde{\ub}_{\mathrm{slip}}$; and by $\bm{\mathrm{K}} \in \mathbb{R}^{3N_X \times 3N_X}$, $\bm{\mathrm{K}}^\dagger \in \mathbb{R}^{3N_X \times 3N_X}$, $\bm{\mathrm{N}} \in \mathbb{R}^{3N_X \times 3N_X}$, and $\bm{\mathrm{F}} \in \mathbb{R}^{3N_X \times 3N_X}$ the matrices that result respectively from the discretization of the operators $\bm{\mathcal{K}}$, $\bm{\mathcal{K}}^\dagger$, $\bm{\mathcal{N}}$, and $\bm{\mathcal{F}}$. The kinematic matrices $\bm{\mathrm{K}}$ and $\bm{\mathrm{K}}^\dagger$ as well as the bending matrix $\bm{\mathrm{F}}$ are derived using classical tools for differentiation and integration that can be found elsewhere \citep{trefethenSpectralMethodsMatlab2000, boydChebyshevFourierSpectral2001} -- we refer the reader to Section 6.1 in Ref. \citep{maxianHydrodynamicsTransientlyCrossLinked2023} for more details on the derivation of these matrices. The hydrodynamic mobility matrix $\bm{\mathrm{N}}$ is similar to its continuum description $\bm{\mathcal{N}}$. To evaluate efficiently the phoretic slip velocity, we rewrite Eq.\eqref{eq: phoretic_slip_velocity} as 

\begin{equation}
    \Tilde{\bm{u}}_{\mathrm{slip}}(s) =  \frac{1}{2}\mathcal{M}(s)\left[2\partial_s \bm{X}(s)\partial_s c^{(0)}(s) -\bm{\mathcal{Q}}(s)\right],
    \label{eq: explicit_phoretic_slip_velocity}
\end{equation}
where $\bm{\mathcal{Q}}$ is given by

\begin{equation}
    \bm{\mathcal{Q}}(s) = -\frac{1}{2}\left[b(s) - 3\right]\rho(s) \mathcal{A}(s) \partial_s^2 \bm{X}(s).
\end{equation}
The above formulation for $\Tilde{\bm{u}}_{\mathrm{slip}}$, Eq.\eqref{eq: explicit_phoretic_slip_velocity}, expresses it explicitly in terms of the centerline position $\bm{X}$, which overcomes issues related to frame orthonormality. We find that this formulation is better conditioned numerically than \eqref{eq: phoretic_slip_velocity}, since it avoids an explicit evaluation of the curvature and the normal vector. Thus, from Eq.\eqref{eq: explicit_phoretic_slip_velocity}, we evaluate the phoretic slip velocity at a collocation point $p$ as 

\begin{equation}
    \discreteSlip_p \equiv \discreteSlip(s_p) = \frac{1}{2}\mathcal{M}_p\left[2\Bigl(\bm{D}^{(3)}_{N_X} \cdot \bm{\mathrm{X}}\Bigr)_p\Bigl(\bm{D}^{(1)}_{N_X} \cdot \bm{\mathrm{c}}^{(0)}\Bigr)_p - \bm{\mathrm{Q}}_p\right],
    \label{eq: dis_phoretic_slip_velocity}
\end{equation}
where $\bm{\mathrm{c}}^{(0)} = \{c^{(0)}(s_p)\}_{p=1}^{N_X} \in \mathbb{R}^{N_X}$ is a vector collecting the leading-order concentration at the collocation points, $\bm{D}^{(d)}_{N_X} \in \mathbb{R}^{dN_X \times dN_X}$ are Chebyshev differentiation matrices \citep{trefethenSpectralMethodsMatlab2000}, and $\bm{\mathrm{Q}} \in \mathbb{R}^{3N_X}$ is the discrete description of $\bm{\mathcal{Q}}$. Analogously to the discretization of the hydrodynamic operator $\bm{\mathcal{N}}$, $\bm{\mathrm{c}}^{(0)}$ and $\bm{\mathrm{Q}}$ are similar to their continuum description.

\subsection{Temporal discretization}
Having discretized the dynamical equations in space, we now turn to their discretization in time. Owing to the stiffness of the problem, i.e., the higher-order derivatives of the bending term induce a fast timescale that limits the region of numerical stability, we use a first-order backward differentiation, where only the bending term $\bm{\mathrm{F}} \cdot \bm{\mathrm{X}}$ is treated implicitly. By doing so, we guarantee numerical stability without the computational cost of a fully implicit method. The resulting discrete saddle point system to solve, at each time step, is thus   

\begin{equation}
    \begin{bmatrix}
    -\bm{\mathrm{N}}_n &\vphantom{\ldots} & \bm{\mathrm{K}}_n \\\\
    \bm{\mathrm{K}}^\dagger_n                               & &\vphantom{\ldots} \bm{0}\\
    \end{bmatrix}
    \cdot
    \begin{bmatrix}
    \bm{\Lambda}_n\\\\
    \bm{\Phi}_n
    \end{bmatrix}
    = 
    \begin{bmatrix}
    \bm{\mathrm{N}}_n \cdot \bm{\mathrm{F}} \cdot \bm{\mathrm{X}}_{n+1} - \discreteSlip_n\\\\
    \bm{0}
    \end{bmatrix}
    \label{eq: discrete_saddle_point_system},
\end{equation}
where $\bm{\Lambda} \in \mathbb{R}^{3N_X}$ and $\bm{\Phi} = \{\bm{\omega}, \bm{\mathrm{U}}_\mathrm{MP}\} \in \mathbb{R}^{3N_X}$ are the discrete descriptions of $\bm{\lambda}(s)$ and $\bm{\phi}(s)=\{\bm{\Omega}(s), \bm{U}(s)|_{s=0}\}$, respectively; and $(\bullet)_n$ denotes the evaluation at the $n$th time step, $n \in \mathbb{N}$. 

Following Ref.\citep{maxianIntegralbasedSpectralMethod2021}, in order to avoid nonlinear solves, we approximate $\bm{\mathrm{X}}_{n+1}$ from the kinematic equation of motion  \eqref{eq:SBT_final_with_lambda} as

\begin{equation}
    \bm{\mathrm{X}}_{n+1} = \bm{\mathrm{X}}_{n} + \Delta t (\bm{\mathrm{K}}_n \cdot \bm{\Phi}_n),
    \label{eq: approx_of_x_next_step}
\end{equation}
where $\Delta t$ is the time-step size. Then, by substituting Eq.\eqref{eq: approx_of_x_next_step} into Eq.\eqref{eq: discrete_saddle_point_system} and rearranging, we get 

\begin{equation}
    \begin{bmatrix}
    -\bm{\mathrm{N}}_n &\vphantom{\ldots} & \widetilde{\bm{\mathrm{K}}}_n\\\\
    \bm{\mathrm{K}}^\dagger_n                               & &\vphantom{\ldots} \bm{0}\\
    \end{bmatrix}
    \cdot
    \begin{bmatrix}
    \bm{\Lambda}_n\\\\
    \bm{\Phi}_n
    \end{bmatrix}
    = 
    \begin{bmatrix}
    \bm{\mathrm{N}}_n \cdot \bm{\mathrm{F}} \cdot \bm{\mathrm{X}}_{n} -  \discreteSlip_n\\\\
    \bm{0}
    \end{bmatrix},
\end{equation}
where $\widetilde{\bm{\mathrm{K}}}_n = \left[\bm{\mathrm{I}} - \Delta t\left(\bm{\mathrm{N}}_n\cdot \bm{\mathrm{F}} \right)\right] \cdot \bm{\mathrm{K}}_n$ with $\bm{\mathrm{I}} \in \mathbb{R}^{3N_X \times 3N_X}$ being the identity matrix. Using the Schur complement, we eliminate $\bm{\Lambda}_n$ in the above system to derive a constrained mobility problem for $\bm{\Phi}_n$,

\begin{equation}
    \bm{\Phi}_n = \consDiscreteHydro_n \cdot \left[\bm{\mathrm{F}} \cdot \bm{\mathrm{X}}_{n} - \left(\bm{\mathrm{N}}_n \right)^{-1} \cdot  \discreteSlip_n\right],
    \label{eq: linear_system_for_discrete_phi_n}
\end{equation}
where $\consDiscreteHydro_n  = \left[\bm{\mathrm{K}}^\dagger_n \cdot \left(\bm{\mathrm{N}}_n \right)^{-1} \cdot  \widetilde{\bm{\mathrm{K}}}_n \right]^{-1} \cdot \bm{\mathrm{K}}^\dagger_n$ is a constraint mobility matrix and can be formed efficiently using pseudo-inverses. Physically, the above expression of $\bm{\Phi}_n$ represents the motion of the filament that results from the competition between the bending and effective phoretic force densities, respectively $\bm{\mathrm{F}}\cdot \bm{\mathrm{X}}_n$ and $-\left(\bm{\mathrm{N}}_n \right)^{-1} \cdot  \discreteSlip_n$.

Once $\bm{\Phi}_n$ is computed, from Eq.\eqref{eq: linear_system_for_discrete_phi_n}, we update the position of the midpoint $\bm{\mathrm{X}}_{\mathrm{MP}}$ as follows: 

\begin{equation}
    \bm{\mathrm{X}}_{\mathrm{MP}, n+1} = \bm{\mathrm{X}}_{\mathrm{MP}, n} + \Delta t \bm{\mathrm{U}}_{\mathrm{MP}, n},
    \label{eq:updateMidpoint}
\end{equation}
and the tangent vectors $\bm{\tau}$ by a rotation around the axes defined by the angles $\bm{\omega}\Delta t$. Let $\bm{\mathrm{q}} = [\cos{(\theta/2)}, \bm{\theta} \sin{(\theta/2)}/\theta]$ be the unit quaternions associated with $\bm{\theta} = \bm{\omega}\Delta t$, where $\theta = \lVert \bm{\theta} \rVert_2$; we obtain $\bm{\tau}_{n+1}$ via   
\begin{equation}
    \bm{\tau}_{n+1} = \bm{\mathrm{R}}(\bm{\mathrm{q}}_n) \cdot \bm{\tau}_{n},
    \label{eq:updateTangent}
\end{equation}
where $\bm{\mathrm{R}}$ is the quaternion-derived rotation matrix \citep{shoemakeAnimatingRotationQuaternion1985}.

After updating the tangent vectors \eqref{eq:updateTangent} and the position of the midpoint \eqref{eq:updateMidpoint}, the configuration of the filament at the next time step, $\bm{\mathrm{X}}_{n+1}$, is then obtained by reconstruction using Eq.\eqref{eq: robot_arm}.

\section{Buckling instability of uniformly active filaments}
\label{section: linear_stability_analysis}
It is well known that a chemical asymmetry is necessary for an isotropically shaped particle such as a rigid sphere to self-propel in a viscous fluid through diffusiophoresis \citep{moranPhoreticSelfPropulsion2017}. In contrast, anisotropically shaped particles such as filaments can take advantage of both their chemical and geometrical asymmetries to self-propel. For instance, a small change in the curvature of a uniformly active straight rod -- which is otherwise immotile since it induces a pure straining flow in the surrounding fluid, i.e., symmetric -- may induce solute concentration gradients on the filament surface and, therefore, lead to a net motion \citep{katsambaChemicallyActiveFilaments2022}. On the other hand, when subject to external compressive forces, a passive flexible filament is susceptible to a buckling instability. In the context of flexible phoretic filaments, the stress induced by phoretic effects, on the filament, is a potential source of mechanical instability. For the sake of intuition, consider a uniformly active filament that is perturbed from its initial straight configuration with a small lateral deflection, $\delta$. Owing to the tangential phoretic slip flow, the filament is subject to an external line tension, $\gamma \sim \eta|AM|a/D$. If this line tension is compressive, then the resulting effective force density -- which is expected to have a destabilizing effect on the filament -- is a product of the line tension with the characteristic curvature, $f_{\mathrm{phoretic}}\sim \gamma \delta/l^2=\eta|AM|a \delta/(Dl^2)$. It can be easily shown that the normal counterpart of the phoretic slip flow induces an effective force density that yields the same scaling. Since the filament is flexible, the restoring elastic force density, $f_{\mathrm{elastic}} \sim B\delta/l^4$, will have a stabilizing effect on it. The filament is therefore susceptible to a mechanical instability that is governed by the balance between the phoretic and elastic force densities, $f_{\mathrm{phoretic}}/f_{\mathrm{elastic}} \sim \eta a l^2 |AM|/(BD)$. This control parameter is nothing other than the elastophoretic number $\alpha$ defined in Section \ref{sec:Chemo_elastohydrodynamic_problem}.  

In the following sections, we investigate -- in more detail -- the stability of a flexible phoretic filament, with uniform activity $\mathcal{A}(s)\equiv \mathcal{A}=\pm1$ and phoretic mobility $\mathcal{M}(s)\equiv \mathcal{M} = \pm1$, against planar perturbations. We assume that the filament radius has a prolate spheroidal profile $\rho(s) = \sqrt{1-s^2}$, so that $b(s)\equiv b = \log(4/\varepsilon^2)$ is uniform along the centerline. The surface concentration field and the resulting slip flow are shown in Fig.\ref{fig:schematic} for $\mathcal{A}=1$, $\mathcal{M}=-1$, and $\varepsilon = 10^{-2}$.

\subsection{Linear stability analysis}
We perform a linear stability analysis of the filament by perturbing it in a plane perpendicular to its major axis. Our approach follows previous works by  Becker and Shelley \citep{beckerInstabilityElasticFilaments2001},  Young and Shelley \citep{youngStretchCoilTransitionTransport2007}, Guglielmini \textit{et al.} \citep{guglielminiBucklingTransitionsElastic2012}, Li \textit{et al.} \citep{liSedimentationFlexibleFilaments2013}, and  Chakrabarti \textit{et al.} \citep{chakrabartiFlexibleFilamentsBuckle2020} on the buckling instability of passive flexible filaments in a viscous fluid.

For small deflections $\epsilon \delta(s)$, with $\epsilon \ll 1$, we use a regular perturbation expansion to write the position of the centerline and the internal line tension as 

\begin{subequations}
    \label{eq:X_and_tension_small_deformation}
\begin{align}
    \Xb(s) &= \Xb_0(s) + \epsilon \Xb_1(s)  + \mathcal{O}(\epsilon^2) \quad \mathrm{and} 
    \label{eq:X_small_deformation}
    \\  
    T(s) &= T_0(s) + \epsilon T_1(s) + \mathcal{O}(\epsilon^2),
    \label{eq:tension_small_deformation}
\end{align}
\end{subequations}
respectively. Note that $\epsilon$ is an arbitrarily small quantity, not to be confused with the filament aspect ratio $\varepsilon$. In its base state (denoted by the subscript $0$), we consider the filament to be straight and aligned along the $x-$axis,
\begin{equation}
    \Xb_0(s) = \Xb_0(s)|_{s=0} + s \xb.
    \label{eq:X0}
\end{equation}
On the other hand, linearizing the inextensibility condition \eqref{eq: inextensibility_condition_summarized} using Eqs.$\eqref{eq:X_small_deformation}$ and \eqref{eq:X0} gives
\begin{align}
    \partial_s\Xb_1(s) \cdot \xb =0.
    \label{eq:linearized_inextensibility}
\end{align}
Accordingly, the $x-$component of $\bm{X}_1$, denoted as $X_{1,x}$, is not a function of $s$, i.e., it is a constant, and therefore, only creates a displacement of the configuration (without deformation) along the $x-$direction. Thus, by translational invariance, we have $X_{1,x}=0$. This implies that the deflection of the filament at the leading order is only along the $y-$direction,
\begin{equation}
    \Xb_1(s) = \delta(s)\yb.
    \label{eq:X1}
\end{equation}
Thus, using Eqs.\eqref{eq:X0} and \eqref{eq:X1}, the position of the centerline \eqref{eq:X_small_deformation} takes the following form: 
\begin{align}
    \Xb(s) = \Xb_0(s)|_{s=0} + s\xb + \epsilon \delta(s) \yb + \mathcal{O}(\epsilon^2),
    \label{eq:X_small_deformation_new}
\end{align}
which gives the tangent and normal vectors as
\begin{subequations}
\begin{align}
    \tb(s) &= \xb + \epsilon \partial_s \delta(s) \yb+ \mathcal{O}(\epsilon^2)\quad \mathrm{and} \\
    \nb(s) &= \yb - \epsilon \partial_s \delta(s) \xb+ \mathcal{O}(\epsilon^2),
\end{align}
    \label{eq:tangent_normal_small_deformation}
\end{subequations}
respectively. 

Let us now determine a regular perturbation expansion for the slip velocity given in Eq.\eqref{eq: phoretic_slip_velocity} for this slightly deformed configuration. For small, smooth deformations, the local curvature can be approximated as $\kappa(s) \approx \epsilon \partial_s^2\delta(s)$. Thus, the normal component of the slip velocity $\tilde{v}$ \eqref{eq:slip_normal} -- being proportional to the curvature -- is intrinsically of $\mathcal{O}(\epsilon)$ at the leading order, for a filament with homogeneous surface chemical properties. For clarity, we write this explicitly by expressing $\tilde{v} \equiv \epsilon \tilde{v}_1\partial_s^2\delta$ and, therefore, Eq.\eqref{eq: phoretic_slip_velocity} takes the form $\tilde{\ub}_\mathrm{slip} = \tilde{u}\tb + \epsilon \tilde{v}_1\partial_s^2\delta\nb$, with 

\begin{subequations}
    \label{eq: slip_uniformly_active_filament}
\begin{align}
    \frac{\tilde{u}(s)}{\mathcal{A}\mathcal{M}} &= \frac{-s b}{2 \sqrt{1-s^2
    }} \quad \mathrm{and} \label{eq:slip_spheroid_tangent}\\
    \frac{\tilde{v}_1(s)}{\mathcal{A}\mathcal{M}} &= \frac{1}{4}(b-3)\sqrt{1-s^2}.
    \label{eq:slip_spheroid_normal}
\end{align}    
\end{subequations}
Thus, using Eq.\eqref{eq:tangent_normal_small_deformation}, $\tilde{\ub}_\mathrm{slip}$ becomes   

\begin{align}
     \tilde{\ub}_\mathrm{slip}(s) = \tilde{u}(s) \xb + \epsilon \left[\tilde{u}(s)\partial_s\delta(s) + \tilde{v}_1(s)\partial_s^2\delta(s)\right] \yb + \mathcal{O}(\epsilon^2).
     \label{eq:slip_small_deformation}
\end{align}
Finally, substituting Eqs.\eqref{eq:X_and_tension_small_deformation}, \eqref{eq:tangent_normal_small_deformation}, and \eqref{eq:slip_small_deformation} in the kinematic equation motion \eqref{eq:SBT_final_with_tension} yields

\begin{align}
    \begin{split}
    \partial_t\Xb_0(s) + \epsilon \partial_t\Xb_1(s) &= \bm{\mathcal{N}} (s) \cdot \biggl[\partial_sT_0(s) \xb + \epsilon\biggl(\partial_sT_1(s)\xb + \partial_s(T_0(s)\partial_s\delta(s))\yb - \frac{1}{\alpha}\partial_s^4\delta(s)\yb\biggr)\biggr] \\ &\quad -
    \left[\tilde{u}(s) \xb + \epsilon (\tilde{u}(s)\partial_s\delta(s) + \tilde{v}_1(s)\partial_s^2\delta(s)) \yb \right]  + \mathcal{O}(\epsilon^2).
    \end{split}
    \label{eq:SBT_small_deformation}
\end{align}
We now analyze Eq.\eqref{eq:SBT_small_deformation} at each asymptotic order.

\subsubsection{$\mathcal{O}(1)$: Straight filament}
\label{sec: leading_order_expansion}
At the leading order, the filament is straight. The curvature is zero throughout, therefore, the slip velocity has only a tangential component. This is evident when collecting terms of $\mathcal{O}(1)$ in Eq.\eqref{eq:SBT_small_deformation}, 

\begin{align}
    \bm{U}_0(s) \equiv \partial_t\Xb_0(s) &= \left[2(b-1) \partial_sT_0(s) - \tilde{u}(s)\right]\xb.
    \label{eq: leading_order_swim_velocity}
\end{align}
From Eq.\eqref{eq:X0}, we observe that the velocity at any point on the centerline of the filament, $\bm{U}_0(s) = \partial_t \Xb_0(s)|_{s=0}\equiv U_0(s)|_{s=0}\xb$, is simply the midpoint velocity -- implying a rigid body motion at this order. Thus, integrating Eq.\eqref{eq: leading_order_swim_velocity} over the length of the filament yields $\bm{U}_0(s)=\bm{0}$, i.e., the filament is stationary at the leading order. This result follows from the free-end boundary condition $T_0(s)|_{s=\pm 1}=0$ and from the fact that $\tilde{u}(s)$ is an odd function (see Eq.\eqref{eq:slip_spheroid_tangent}), i.e., its integral vanishes over the length of the filament. The tangential slip flow, however, drives a straining flow in the surrounding fluid, which induces a compressive or tensile stress on the filament. This stress gives rise to an internal elastic restoring force (internal line tension) that acts against the external tangential solicitation. The leading-order internal line tension $T_0$ is obtained by substituting Eq.\eqref{eq:slip_spheroid_tangent} in Eq.\eqref{eq: leading_order_swim_velocity} and solving,

\begin{equation}
    \frac{T_0(s)}{\mathcal{A}\mathcal{M}} = \frac{b}{4(b-1)} \sqrt{1-s^2}.
    \label{eq:T0}
\end{equation}
Note that since $b > 1$, the signed quantity $\mathcal{A}\mathcal{M}=1$ implies that the tangential stress induced by phoretic effects is tensile ($T_0\geq0$) whereas when $\mathcal{A}\mathcal{M}=-1$, it is compressive ($T_0\leq0$) -- buckling is therefore susceptible to occur in the latter case.

\subsubsection{$\mathcal{O}(\epsilon)$: Small deformation}
\label{section: small_deformation_lsa}
Collecting terms of $\mathcal{O}(\epsilon)$ in Eq.\eqref{eq:SBT_small_deformation}, we get

\begin{figure}[t]
\centering
    \includegraphics[width=0.95\textwidth]{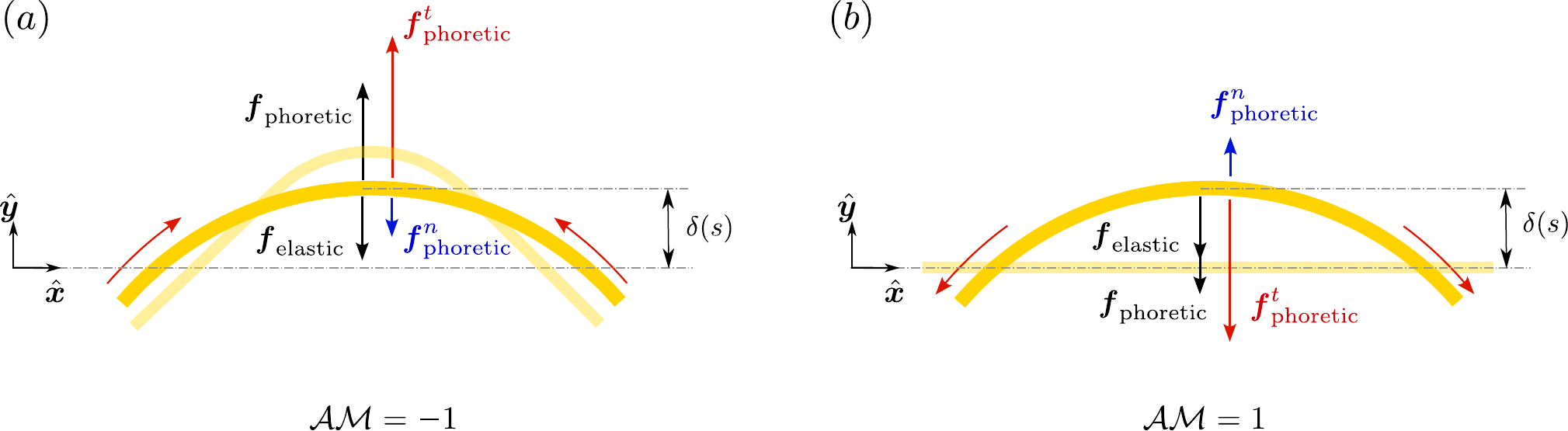}
\caption{Illustration of the instability mechanism in an autophoretic filament with uniform surface activity $\mathcal{A}$ and phoretic mobility $\mathcal{M}$. The tangential phoretic slip induces an external line tension on the filament, shown by the curved red arrows. Owing to the small deflection $\delta$, this external line tension results in an effective normal force, $\bm{f}^t_\mathrm{phoretic}$. On the other hand, the normal phoretic slip also results in an effective normal force, $\boldsymbol{f}^n_\mathrm{phoretic}$. These phoretic forces oppose each other, so the filament is subject to the resulting net force, $\fb_\mathrm{phoretic} = \boldsymbol{f}^t_\mathrm{phoretic} + \boldsymbol{f}^n_\mathrm{phoretic}$. The elastic response $\fb_\mathrm{elastic}$ always acts to restore the straight configuration of the filament, while $\fb_\mathrm{phoretic}$ can act to either restore or destabilize the filament depending on the surface chemical properties. $(a)$ $\mathcal{A}\mathcal{M}=-1$, the net phoretic force amplifies the deflection, leading to a buckling instability. $(b)$ $\mathcal{A}\mathcal{M}=1$, both the net phoretic force and the elastic force restore the filament to its straight configuration. For notation convenience, we have represented forces using force density symbols.} 
\label{fig: illustrationOfBuckling}
\end{figure}

\begin{align}
    \partial_t\Xb_1(s) = (b+1) \partial_sT_1(s)\xb  + \biggl[(b-3)\partial_sT_0(s)\partial_s\delta(s) + (b+1) \partial_s(T_0(s)\partial_s\delta(s)) \biggr. \nonumber\\
    \biggl. - \frac{(b+1)}{\alpha} \partial_s^4\delta(s) - (\tilde{u}(s)\partial_s\delta(s) + \tilde{v}_1(s)\partial_s^2\delta(s))\biggr]\yb.
    \label{eq: first_order_swim_velocity}
\end{align}
Using Eq.\eqref{eq:X1} and dotting Eq.\eqref{eq: first_order_swim_velocity} with the unit vector $\xb$, we get $\partial_sT_1(s)=0$. The latter implies that the first-order internal line tension is uniform along the filament centerline and, therefore, must vanish since the filament has free ends, $T_1(s)=0$. On the other hand, dotting Eq.\eqref{eq: first_order_swim_velocity} with the unit vector $\yb$ and, using Eqs.\eqref{eq: slip_uniformly_active_filament} and \eqref{eq:T0}, we get the governing equation for the growth rate of the perturbation 

\begin{align}
    \partial_t\delta(s) = \frac{\mathcal{A}\mathcal{M}}{4} \sqrt{1-s^2}\left(\frac{5b-3}{b-1}\right)\partial_s^2\delta(s) - \frac{(b+1)}{\alpha} \partial_s^4\delta(s),
    \label{eq:deltadot}
\end{align}
subject to the free-end boundary conditions $\partial^2_s\delta(s)|_{s=\pm 1} = 0$ and $\partial^3_s \delta(s)|_{s=\pm 1} =0$. The two terms on the right-hand side of Eq.\eqref{eq:deltadot} result from phoretic and bending effects, respectively.

The curvature dependence of the phoretic term in Eq.\eqref{eq:deltadot} results from (i) the leading-order internal line tension $T_0$ -- induced by the tangential component of the slip $\tilde{u}$ -- and (ii) the normal component of the slip $\Tilde{v}_1 \partial^2_s\delta$,

\begin{equation}
    \frac{\sqrt{1-s^2}}{4}\left(\frac{5b-3}{b-1}\right)\partial_s^2\delta(s) = (b+1)\frac{T_0(s)}{\mathcal{A}\mathcal{M}}\partial_s^2\delta(s) - \frac{\Tilde{v}_1}{\mathcal{A}\mathcal{M}} \partial^2_s\delta(s).
    \label{eq: phoretic_term_growth_rate}
\end{equation}
While the curvature dependence of the first term on the right-hand side of Eq.\eqref{eq: phoretic_term_growth_rate} is the hallmark of classical buckling instability, i.e., the resulting transverse force on a slightly bent rod under tangential compressive stresses is proportional to the curvature \citep{landauTheoryElasticityVolume1986}, that of the second term arises purely from phorectic effects since the normal component of the slip inherently scales with the curvature, thus introducing a distinct source of mechanical instability. But do these two effects reinforce or oppose each other? It is clear from Eqs.\eqref{eq:slip_spheroid_normal} and \eqref{eq:T0} that both $\Tilde{v}_1/\mathcal{A}\mathcal{M}$ and $T_0/\mathcal{A}\mathcal{M}$ are positive. Therefore, since these terms appear with opposite signs in the governing equation for the growth rate, they counteract each other: If one promotes destabilizing effects, then the other promotes the opposite. The net phoretic effect is therefore governed by the dominant contribution. Since $b > 1$, their difference in Eq.\eqref{eq: phoretic_term_growth_rate} is positive, implying that the effect of the tangential component dominates. Tangential stresses can lead to a buckling instability, requiring that the condition $T_0 \leq 0$ be satisfied -- which corresponds to $\mathcal{A}\mathcal{M}=-1$, as shown in Section \ref{sec: leading_order_expansion}. This implies that filaments with positive (respectively, negative) phoretic mobility $\mathcal{M}$ will experience a buckling instability only if they have a negative (respectively, positive) activity $\mathcal{A}$ which corresponds to an absorption (respectively, emission) of the solute on the surface of the filament. Figure \ref{fig: illustrationOfBuckling} shows an illustration of the instability mechanism. Nonetheless, with different surface chemical properties, the effect of the normal slip may lead to an instability via the alternate route.

\subsection{Eigenvalue problem}
We seek exponentially growing function for the perturbation, $\delta(s) = \sum_n \varphi_n(s) e^{\sigma_n t}$. Accordingly, Eq.\eqref{eq:deltadot} yields the following eigenvalue problem: 
\begin{equation}
    \sigma_n \varphi_n(s) = \frac{\mathcal{A}\mathcal{M}}{4} \sqrt{1-s^2}\left(\frac{5b-3}{b-1}\right) \partial_s^2\varphi_n(s) - \frac{(b+1)}{\alpha} \partial_s^4\varphi_n(s),
    \label{eq:EVP}
\end{equation}
for the mode configurations $\varphi_n(s)$ and the corresponding eigenvalues $\sigma_n$. 

\begin{figure}
\centering
    \includegraphics[width=0.95\textwidth]{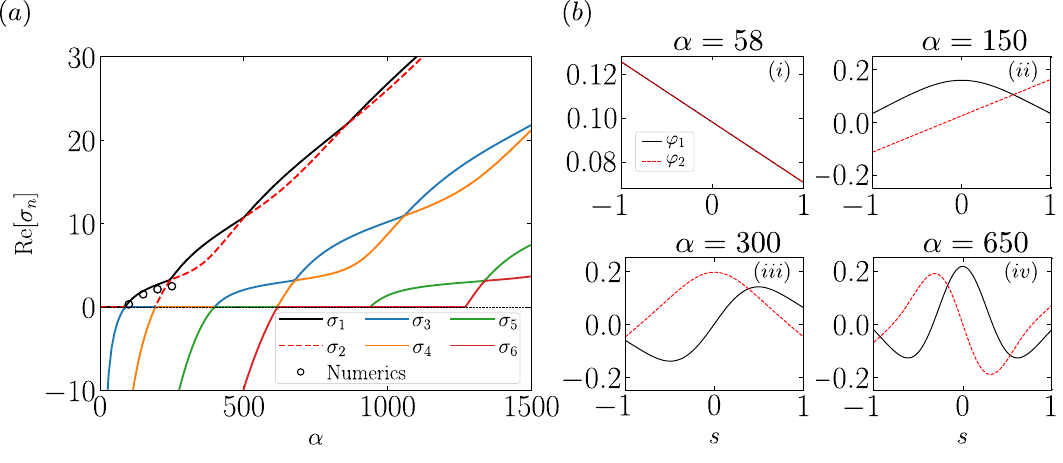}
\caption{$(a)$ Real parts of the first six eigenvalues (growth rates) $\sigma_n$ as function of the elastophoretic number $\alpha$. The lines and circles correspond to theoretical predictions and numerical simulations, respectively. $(b)$ Eigenfunctions of the two most unstable modes, $\varphi_1$ (solid line) and $\varphi_2$ (dashed line), for different values of the elastophoretic number $\alpha$. Parameter values are $\mathcal{A}\mathcal{M}=-1$ and $\varepsilon=10^{-2}$.}
\label{fig:EigenValueSpectrum_and_EigenFunctions}
\end{figure}

We used a pseudospectral method \citep{trefethenSpectralMethodsMatlab2000, boydChebyshevFourierSpectral2001} to solve Eq.\eqref{eq:EVP} along with the boundary conditions $\partial_s^2\varphi_n(s)|_{s=\pm1}=0$ and $\partial_s^3\varphi_n(s)|_{s=\pm1}=0$. For $\mathcal{AM}=1$, all the eigenvalues are found to be negative for any value of the elastophoretic number $\alpha$, implying that the perturbed filament relaxes back to its straight configuration. 
In Fig.\ref{fig:EigenValueSpectrum_and_EigenFunctions}$(a)$, we show, for $\mathcal{A}\mathcal{M}=-1$ and $\varepsilon=10^{-2}$, the real parts of the first six eigenvalues for increasing values of $\alpha$. We find that the first two eigenmodes, $\sigma_1$ and $\sigma_2$, become unstable beyond $\alpha_{c1} \approx 87$ and $\alpha_{c2} \approx 193$, respectively. These two modes remain the most unstable ones as $\alpha$ increases, i.e., when the filament becomes more flexible. In Fig.\ref{fig:EigenValueSpectrum_and_EigenFunctions}$(b)$, we show the corresponding eigenfunctions $\varphi_1$ and $\varphi_2$ -- physically describing the respective filament conformations -- for various values of $\alpha$. Below the buckling threshold, $\alpha < \alpha_{c1}$, all the modes are stable and, therefore, the filament remains straight (Fig.\ref{fig:EigenValueSpectrum_and_EigenFunctions}$(b)$, panel $(i)$). For $\alpha=150$ (Fig.\ref{fig:EigenValueSpectrum_and_EigenFunctions}$(b)$, panel $(ii)$), the first mode is unstable while the second remains stable. The filament conformation, $\varphi_1$, has a nonzero curvature and resembles a “U" shape. For $\alpha > \alpha_{c2}$, more complex conformations with larger wavenumbers -- such as ``S" and ``W" shapes -- arise  (Fig.\ref{fig:EigenValueSpectrum_and_EigenFunctions}$(b)$, panels $(iii)$ and $(iv)$). Note that in this case, $\varphi_1$ and $\varphi_2$ always come in an odd-even pair; an analogous result was found for the buckling instability of a passive filament under a compressional flow \citep{chakrabartiFlexibleFilamentsBuckle2020}.

\section{Nonlinear dynamics of uniformly active filaments}
\label{section:nonlinearDynamics}
The linear stability analysis performed in Section \ref{section: linear_stability_analysis} provides some insights into the spontaneous filament dynamics at the early stage of buckling. The buckled configurations are geometrically asymmetric, resulting in self-propulsion of the filament. In this section, we investigate the nonlinear dynamics of the filament and characterize its swimming modes in terms of the elastophoretic number. To do so, we solve the fully nonlinear chemoelastohydrodynamic problem introduced in Section \ref{section: chemo_elasto_hydro_problem} using the numerical methods described in Section \ref{section: numerical_methods}.

\subsection{Validation of the numerical approach}
\label{section:validation_of_the_numerical_approach}
We first validate our numerical approach against theoretical predictions from the linear stability analysis. Following Ref.\citep{guglielminiBucklingTransitionsElastic2012}, we expand the filament deflection $\delta(s)= \bm{\hat{n}}(s) \cdot [\bm{X}(s) - \bm{X}(s)|_{s=0}]$, obtained from numerical simulations, in a basis formed by the eigenfunctions $\psi_n(s)$ of the biharmonic bending operator $\bm{\mathcal{F}}$ that satisfy the free-end boundary conditions \citep{wigginsTrappingWigglingElastohydrodynamics1998, loughSelfbucklingSelfwrithingSemiflexible2023}. Thus, $\delta(s) =\sum_n a_n(t)\psi_n(s)$ with $a_n = \langle \psi_n^{\dagger}, \delta\rangle/\langle \psi_n^{\dagger}, \psi_n\rangle$, where $\langle \bullet, \bullet \rangle$ denotes the inner product and $\psi_n^{\dagger}$ are the eigenfunctions of the adjoint problem. The latter are equivalent to $\psi_n$, as the bending operator is Hermitian. Having the coefficients $a_n$, we compute the growth rates $\mathrm{Re}[\sigma_n] = \mathrm{\partial}_t a_n(t)|_{t=0}/a_n(0)$, with $a(0)=10^{-8}$. We find a good agreement between the results from numerical simulations and linear stability analysis near the threshold, see Fig.\ref{fig:EigenValueSpectrum_and_EigenFunctions}$(a)$. Quantitatively, the former differ relatively from the latter by $18\%$ to $37\%$ for $11 \leq \Delta \alpha \leq  161$, where $\Delta \alpha$ is the offset from the threshold. The difference stems from (i) nonlinearities, which become significant far from the threshold; and (ii) the numerical error due to the nearly singular behavior of the tangential component of the phoretic slip velocity, $\Tilde{u}$, in the vicinity of the endpoints\footnote{The use of a Chebyshev basis of the first kind, i.e., that excludes the endpoints ($s=\pm1$), prevents $\Tilde{u}$ from exhibiting singular behavior.} -- see Eq.\eqref{eq:slip_spheroid_tangent}. In Appendix \ref{appendix: convergence_study}, we conduct a convergence study to verify the accuracy of our numerical scheme.

\subsection{Results and discussion}
\begin{figure}
    \centering
    \includegraphics[width=0.835\linewidth]{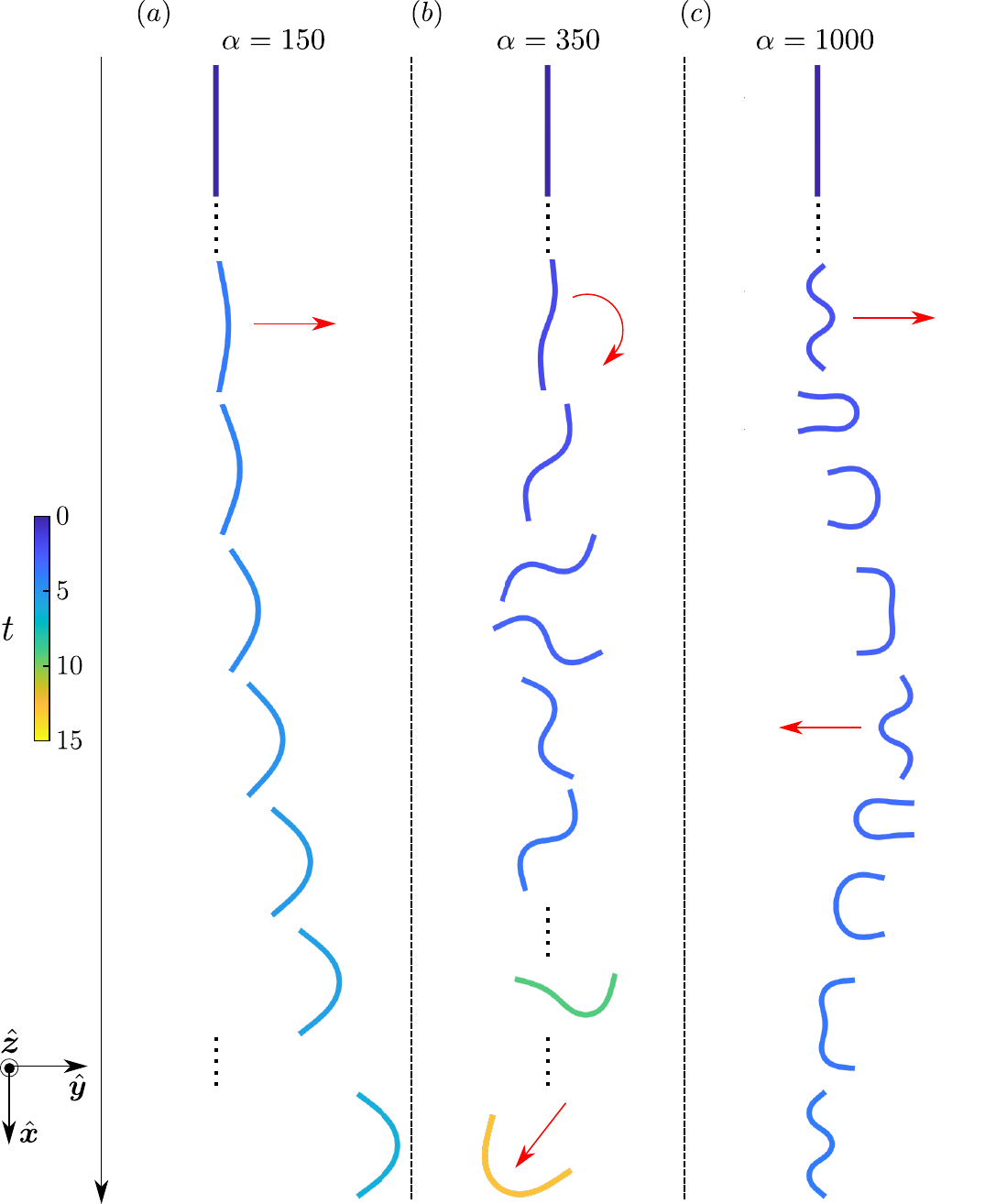}
    \caption{Chronophotographies of the buckling-induced self-propulsion of an autophoretic filament for different values of the elastophoretic number $\alpha$. The filament is initially perturbed from its straight configuration with a small deflection, and the amplitude of the perturbation grows over time -- leading to self-propulsion. $(a)$ Steady “U" shape for $\alpha=150$. $(b)$ Metastable “S" shape at the transient regime which evolves into a steady “U" shape for $\alpha=350$. $(c)$ Oscillating motion for $\alpha=1000$. The black and red arrows denote the time evolution and swimming direction, respectively. Parameter values are $\mathcal{AM}=-1$ and $\varepsilon=10^{-2}$.}
    \label{fig: chronophotographies_nonlinear_dynamics}
\end{figure}

We perform numerical simulations for $\alpha \geq \alpha_{c1}$, $\mathcal{A}\mathcal{M}=-1$, and $\varepsilon=10^{-2}$. For each simulation, a straight filament is initially perturbed either with the first even, $a_1\psi_1$, or the first odd, $a_2\psi_2$, biharmonic eigenfunction of the bending operator, where $a_1 = a_2 =  10^{-5}$. The considered time interval is $0 \leq t \leq 15$ and, for better accuracy (see Appendix \ref{appendix: convergence_study}), we set the spatial resolution and the time-step size to $N_X=101$ and $\Delta t=10^{-4}$, respectively. Figure \ref{fig: chronophotographies_nonlinear_dynamics} shows the chronophotographies, i.e., time sequences, of the filament dynamics for $\alpha = 150$ (panel $(a)$), $\alpha = 350$ (panel $(b)$), and $\alpha = 1000$ (panel $(c)$). The amplitude of the perturbation is found to increase over time, thereby enhancing the shape asymmetry necessary for the filament to self-propel. Near the threshold, $\alpha = 150$, the filament adopts a steady “U" shape after a transient regime (see also movies 1a and 1b in the Supplemental Material \citep{SSM}); whereas it exhibits an oscillating motion in the highly flexible regime, $\alpha=1000$ (see also movies 3a and 3b in the Supplemental Material \citep{SSM}). For an intermediate value of the elastophoretic number, $\alpha=350$, when initialized with $a_2\psi_2$, the filament adopts a rotating “S" shape (see also movies 2a and 2b in the Supplemental Material \citep{SSM}) which turns out to be a metastable configuration as it becomes a steady “U" at long times. 

\begin{figure}[htb!]
    \centering
    \includegraphics[width=0.95\linewidth]{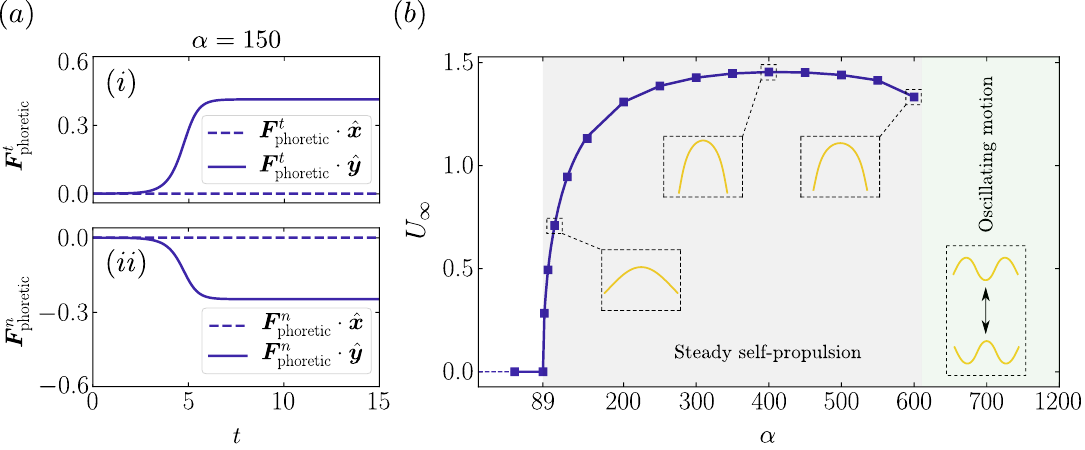}
    \caption{$(a)$ Time evolution of the effective forces induced by the tangential (panel $(i)$) and normal (panel $(ii)$) phoretic slip flows for $\alpha=150$. The net force $\bm{F}_\mathrm{phoretic}= \bm{F}_\mathrm{phoretic}^t + \bm{F}_\mathrm{phoretic}^n$ -- which defines the swimming direction -- is dominated by the tangential contribution. $(b)$ Evolution of the steady swimming velocity $U_\infty$ with the elastophoretic number $\alpha$. The ranges of the “U”-shaped self-propulsion and oscillating regimes are shown by the gray- and green-shaded zones, respectively. The insets show the configurations of the filament at the indicated values of $\alpha$. Parameter values are $\mathcal{AM}=-1$ and $\varepsilon=10^{-2}$.}
    \label{fig: phoretic_forces_and_evolution_steady_swimming_velocity_with_alpha}
\end{figure}

An understanding of the swimming direction can be inferred by examining the effects of the tangential and normal phoretic slip flows, which result in an effective phoretic force density, $-\bm{\mathrm{N}}^{-1}\cdot\discreteSlip$ (see Eq.\eqref{eq: linear_system_for_discrete_phi_n}). We obtain the resulting force, $\bm{F}_\mathrm{phoretic} \in \mathbb{R}^{3}$, by integrating the latter over the filament length,

\begin{equation}
    \bm{F}_\mathrm{phoretic} = -\bm{W}\cdot \bm{\mathrm{N}}^{-1} \cdot \discreteSlip,
\end{equation}
where $\bm{W} = \mathrm{diag}(\bm{w}^T, \bm{w}^T, \bm{w}^T) \in \mathbb{R}^{3 \times 3N_X}$ with $\bm{w} \in \mathbb{R}^{3N_X}$ being a column vector of integration weights. Figure \ref{fig: phoretic_forces_and_evolution_steady_swimming_velocity_with_alpha}$(a)$ shows the time evolution of the tangential and normal components of $\bm{F}_\mathrm{phoretic}$ for $\alpha=150$, respectively $\bm{F}_\mathrm{phoretic}^t$ and $\bm{F}_\mathrm{phoretic}^n$. We see that the direction of motion (see the red arrow in Fig.\ref{fig: chronophotographies_nonlinear_dynamics}$(a)$) is clearly governed by the competition between these forces, which is dominated by the tangential contribution, therefore the tangential slip. This finding agrees with our theoretical predictions -- in the small-deformation limit --, namely, the growth rate of the perturbation is driven by the tangential phoretic slip flow (see Section \ref{section: small_deformation_lsa}). Accordingly, the direction of rotation (respectively, translation) of the “S" (respectively, “W") shape, indicated by the red arrow in Fig.\ref{fig: chronophotographies_nonlinear_dynamics}$(b)$ (respectively, Fig.\ref{fig: chronophotographies_nonlinear_dynamics}$(c)$), can be inferred from that of a “U" by forming a “S" (respectively, “W") by joining two “U" (respectively, three “U") shapes.

Overall, our simulations reveal that, regardless of the initial perturbation, the long-time dynamics of the filament lead to one of two swimming regimes: steady “U"-shaped self-propulsion or oscillating motion. The ranges, in terms of the elastophoretic number $\alpha$, of these swimming regimes are depicted in Fig.\ref{fig: phoretic_forces_and_evolution_steady_swimming_velocity_with_alpha}$(b)$, which shows the evolution of the steady swimming velocity $U_\infty$ with $\alpha$. We find that the lower bound, i.e., the threshold, of the first regime (in gray in Fig.\ref{fig: phoretic_forces_and_evolution_steady_swimming_velocity_with_alpha}$(b)$), steady self-propulsion, is about $\alpha_c \approx 89$, which agrees with our theoretical prediction, $\alpha_{c1} \approx 87$; and, the lower bound of the second regime (in green in Fig.\ref{fig: phoretic_forces_and_evolution_steady_swimming_velocity_with_alpha}$(b)$), oscillating motion, about $\alpha \approx 605$. Although we have reported the upper bound of the latter regime as $\alpha \approx 1200$ in Fig.\ref{fig: phoretic_forces_and_evolution_steady_swimming_velocity_with_alpha}$(b)$, we would like to point out to the reader that the exact value may be larger than this. Indeed, simulations for very high values of $\alpha$ are subject to numerical instabilities, since the bending energy is close to zero; the filament becomes unstable under any stress and exhibits large changes in the curvature, making simulations in this regime very challenging. Moreover, in this regime, the assumptions underlying local SBT and SPT break down, and nonlocal contributions become significant. We therefore do not attempt to resolve this limit quantitatively; nonetheless, we expect the filament dynamics to become chaotic.

Now that the swimming regimes have been identified and described, we characterize, in the following, the resulting kinematics and deformation of the filament in each of them.

\subsubsection{Steady self-propulsion}
As shown in Fig.\ref{fig: phoretic_forces_and_evolution_steady_swimming_velocity_with_alpha}$(b)$, the steady swimming velocity $U_\infty$ increases beyond the bifurcation at $\alpha_c \approx 89$, reaches a maximum at $\alpha\approx 400$, and then decreases with increasing $\alpha$. This supercritical behavior may not seem intuitive at first glance, but it can be rationalized using scaling arguments.

At the steady state, the filament undergoes a rigid body motion. Therefore, its swimming velocity solely depends on the effective phoretic force density $f_\mathrm{phoretic}$ (see Section \ref{section: linear_stability_analysis}),

\begin{equation}
    U_\infty \sim \frac{f_\mathrm{phoretic}}{\eta} \equiv \frac{|AM|a \delta_\infty}{Dl^2},
\end{equation}
where $\delta_\infty = \mathrm{max}\{\hat{\bm{n}}(s)\cdot [\bm{X}_\infty(s) - \bm{X}_\infty(s)|_{s=0}]\}$ is the maximal deflection of the filament, with $\bm{X}_\infty$ being the position of its centerline at the steady state. Noting that $|AM|a/(Dl) = V$ is the characteristic phoretic slip velocity, the above expression yields $U_\infty \propto \delta_\infty$. We see that the steady swimming velocity is a monotonic function of the maximal deflection. This implies that the behavior of the former can be inferred from the latter. Indeed, as the maximal deflection increases (respectively, decreases), the viscous drag decreases (respectively, increases) since larger (respectively, smaller) parts of the filament are aligned in the swimming direction.    

\begin{figure}[ht]
    \centering
    \includegraphics[width=0.95\linewidth]{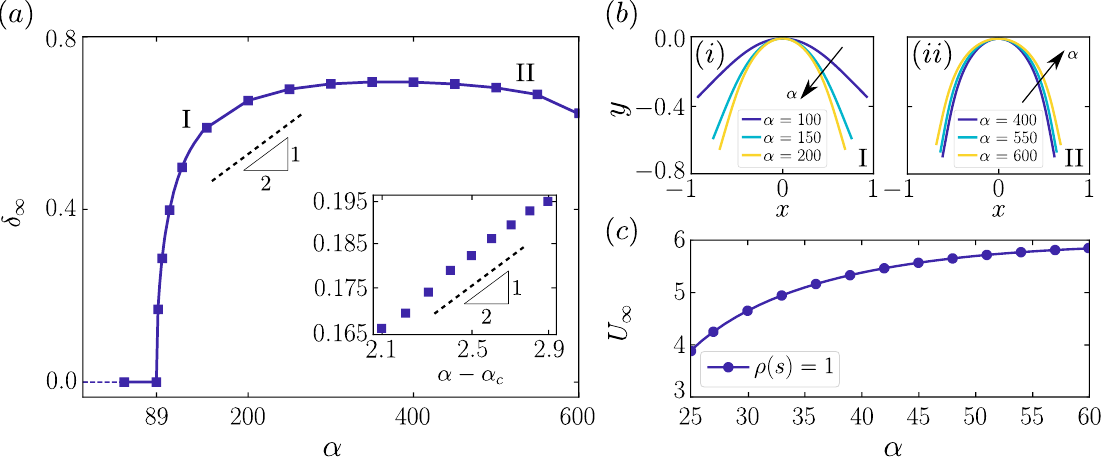}
    \caption{Steady self-propulsion. $(a)$ Evolution of the maximal deflection at the steady state $\delta_\infty$ with the elastophoretic number $\alpha$. The inset shows the behavior of $\delta_\infty$ near the threshold, $\delta_\infty \propto \sqrt{\alpha - \alpha_c}$, which is the canonical scaling of a pitchfork bifurcation. $(b)$ Steady-state configurations of the filament for different values of $\alpha$: $(i)$ $\delta_\infty$ increases with increasing $\alpha$ (corresponding to zone I in $(a)$), and $(ii)$ $\delta_\infty$ decreases with increasing $\alpha$ (corresponding to zone II in $(a)$). $(c)$ Evolution of the steady swimming velocity $U_\infty$ with the elastophoretic number $\alpha$ for a cylindrical profile of the cross-sectional radius, $\rho(s)=1$. Parameter values are $\mathcal{AM}=-1$ and $\varepsilon=10^{-2}$.}
    \label{fig: steady_self_propulsion_regime}
\end{figure}

Figure \ref{fig: steady_self_propulsion_regime}$(a)$ shows the evolution of the maximal deflection $\delta_\infty$ with the elastophoretic number $\alpha$ at the steady state. Near the threshold, $\alpha_c$, the maximal deflection is proportional to $\sqrt{\alpha - \alpha_c}$, see the inset. This finding implies

\begin{equation}
    U_\infty \propto \sqrt{\alpha - \alpha_c},
\end{equation}
and therefore agrees with the expected canonical scaling of a pitchfork bifurcation \citep{glendinningStabilityInstabilityChaos1994, charruInstabilitesHydrodynamiques2007}. Although the mechanisms that lead to the instability are different, the obtained scaling has also been observed in several studies \citep{huChaoticSwimmingPhoretic2019, liSwimmingDynamicsSelfpropelled2022, zhuSelfpropulsionEllipticalPhoretic2023, cobosSpontaneousLocomotionSymmetric2024}; however it differs from the case of isotropically shaped autophoretic particles investigated by Michelin \textit{et al.} \citep{michelinSpontaneousAutophoreticMotion2013}, where they found the swimming velocity to be proportional to  $\mathrm{Pe} - \mathrm{Pe}_\mathrm{c}$ with $\mathrm{Pe}$ being the Péclet number.

When $\delta_\infty$ increases, the steady shape of the filament evolves from a less to a more pronounced “U", see panel $(i)$ in Fig.\ref{fig: steady_self_propulsion_regime}$(b)$. Further away from the threshold, $92 \lesssim \alpha \lesssim 350$, we find $\delta_\infty$ to be proportional to $\alpha^{1/2}$. This scaling results from the competition between the stress induced by the phoretic slip flow and the elastic restoring force. Indeed, by balancing the bending energy $\mathcal{E}_\delta \sim B \delta_\infty^2 / l^3$ and the work done by the effective phoretic force $J \sim \eta |AM|al/D$, we obtain $\delta_\infty \propto \alpha^{1/2}$. The swimming velocity is thus 

\begin{equation}
    U_\infty \propto  \alpha^{1/2}. 
\end{equation}
Although not an instability, this behavior is reminiscent of the sedimentation of a flexible filament in a quiescent viscous fluid. Marchetti \textit{et al.} \citep{marchettiDeformationFlexibleFiber2018} found a similar evolution of the shape from a less to a more pronounced “U". In the reconfiguration regime, they found a similar scaling for the settling velocity in terms of the elastoviscous number.  

Once $\delta_\infty$ has reached its maximum value ($\approx 0.7$), it starts to decrease with increasing $\alpha$, see panel $(ii)$ in Fig.\ref{fig: steady_self_propulsion_regime}$(b)$. As a result, the swimming velocity is hindered in $400 \lesssim \alpha \lesssim 600$. This decrease results from a reduction in the magnitude of the phoretic forcing when the filament has reached its limit curvature, which corresponds to the maximum value of $\delta_\infty$. We identified the profile of the cross-sectional radius as a source of this reduction. Indeed, by using a cylindrical profile $\rho(s)=1$ , $\delta_\infty$ strictly increases as $\alpha$ increases (Fig.\ref{fig: steady_self_propulsion_regime}$(c)$). This finding implies that the profile of the cross-sectional radius induces restoring phoretic stresses, thereby reducing the net phoretic forcing \citep{michelinGeometricTuningSelfpropulsion2017}. 

\subsubsection{Oscillating motion}
For large values of the elastophoretic number ($605 \lesssim \alpha \lesssim 1200$), owing to highly nonlinear effects, the filament exhibits an oscillating motion. A similar regime has also been observed in the buckling of elongated cells under their flagellar activity \citep{loughSelfbucklingSelfwrithingSemiflexible2023} and of microtubules under the action of molecular motors \citep{manMorphologicalTransitionsAxiallydriven2019}.  

\begin{figure}[ht]
    \centering
    \includegraphics[width=0.95\linewidth]{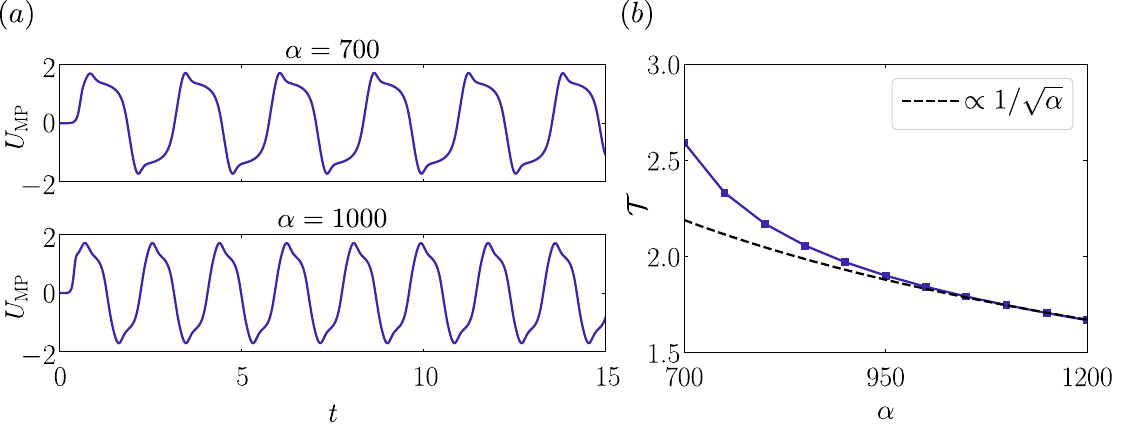}
    \caption{Oscillating motion. $(a)$ Time evolution of the swimming velocity of the filament at its midpoint, $U_\mathrm{MP} = U(s)|_{s=0}$, for different values of the elastophoretic number $\alpha$. $(b)$ Evolution of the oscillation period $\mathcal{T}$ with the elastophoretic number $\alpha$. $\mathcal{T}$ scales as $\alpha^{-1/2}$ for highly flexible filaments. Parameter values are $\mathcal{AM}=-1$ and $\varepsilon=10^{-2}$.}
    \label{fig: oscillatingMotion}
\end{figure}

In order to characterize the oscillating motion of the filament, we track the swimming velocity at its midpoint, $U_\mathrm{MP} = U(s)|_{s=0}$, over time for different values of $\alpha$ (see Fig.\ref{fig: oscillatingMotion}$(a)$). We observe that $U_\mathrm{MP}$ oscillates with a constant amplitude regardless of $\alpha$ but with a period that is sensitive to the latter. This implies that the conformations of the filament -- over time (as shown in Fig.\ref{fig: chronophotographies_nonlinear_dynamics}$(c)$) -- are independent of $\alpha$, but the amplitude of the displacement does depend on $\alpha$. 

The dependence of the oscillation period $\mathcal{T}$ on $\alpha$ is shown in Fig.\ref{fig: oscillatingMotion}$(b)$. $\mathcal{T}$ decreases at a rapid rate for smaller values of $\alpha$ and then slows down for larger values. In the latter, the oscillation period scales as $\alpha^{-1/2}$ -- which can be rationalized using scaling arguments. Let $\mathcal{T}$ be a function of the amplitude of the displacement $d$ and the characteristic swimming velocity $U_\mathrm{MP}$,

\begin{equation}
    \mathcal{T} \sim \frac{d}{U_\mathrm{MP}}.
\end{equation}
Since the amplitude of $U_\mathrm{MP}$ is constant over time and independent of $\alpha$, a suitable choice of its scale is the characteristic phoretic slip velocity, $U_\mathrm{MP} \sim V$. On the other hand, $d$ is controlled by the characteristic bending energy $\mathcal{E} \sim B/l$ and the work done by the effective phoretic force $J \sim \eta |AM|al/D$, $d = \mathrm{function}(\mathcal{E}, J)$. Thus, a simple scaling analysis gives $d/l \sim \alpha^{-1/2}$ and then implies

\begin{equation}
    \frac{\mathcal{T}}{\tau_\mathrm{phoretic}} \sim \alpha^{-1/2} \Rightarrow \mathcal{T} \propto \alpha^{-1/2}. 
\end{equation}

\section{Conclusions}
\label{section:conclusions}
In this paper, we have identified and characterized a novel route to self-propulsion in autophoretic colloids, involving symmetry breaking from a buckling instability. The mechanism differs fundamentally from the ones previously studied, such as surface chemical patterning and advective instability, in that it relies on the mechanical compliance of the particle. We illustrated this concept by studying the simplest example: an active elastic filament with uniform surface chemical properties. Thus, flexibility enables self-propulsion of filaments without requiring any heterogeneous patterning as in autocatalytic Janus colloids. This work, therefore, challenges the traditional view of buckling as an undesirable failure mode.

At the heart of the mechanism is the competition between elastic restoring forces and phoretic destabilizing stresses. An interesting subtlety is the interplay between the effects of the tangential and normal (azimuthal) components of the phoretic slip velocity, whose relative strengths depend on the chemical patterning of the surface. We have shown here that for filaments having uniform chemical activity $A$ and phoretic mobility $M$, the stresses from the tangential component dominate, and the filament is susceptible to a buckling instability if $AM<0$. This finding contrasts with the case of advective instability observed in uniformly active colloids, where $AM>0$ is the necessary condition \citep{michelinSpontaneousAutophoreticMotion2013,zhuSelfpropulsionEllipticalPhoretic2023}.
However, for a surface chemical pattern in which the effect of the normal component dominates, the necessary condition for the mechanical instability of the filament reverses. 

The elongated nature of the flexible filament allows it to achieve multiple swimming modes which are functionally different: stationary pumping when straight, “U"-shaped steady translation, metastable “S"-shaped rotation, and periodic oscillations. Experiments of active rods having these fixed shapes indeed exhibit similar translational and rotational motion \citep{sharanFundamentalModesSwimming2021}. Since all surface slip flows, irrespective of their origin, have the potential to destabilize an elastic filament, we expect our analysis to equally apply to other similar interfacial phenomena such as electrophoresis or thermophoresis. Even some of the generic active filament models described in the introduction (Section \ref{sec:introduction}) report qualitatively similar dynamics, including spontaneous symmetry breaking with “U"- and “S"-shaped self-propelling modes, flapping, and self-writhing. While the onset of motion depends on the geometric asymmetry of the buckled configurations, the long-time configuration and swimming behavior are governed by the nonlinear coupling with the surrounding fluid.

In this study, we only considered planar deformations. Recent numerical simulations by Butler \textit{et al.} \citep{butlerElastohydrodynamicsThreedimensionalChemically2026} -- that include out-of-plane bending and twisting in the model -- report three-dimensional deformations for highly flexible filaments with nonhomogeneous surface chemical properties. For a symmetric activity profile, $\mathcal{A}(s)=\sqrt{1-s^2}$, they observe planar configurations when $125 \lesssim \alpha < 1344$ and a transition to dynamic out-of-plane configurations when $\alpha \gtrsim 1344$. In the latter regime, they observe rich dynamics, including helical, pinwheeling, ballistic-like, and diffusive-like trajectories, as $\alpha$ increases. While it is certainly important to explore the parameter space of three-dimensional filament dynamics, we have shown here that even planar conformations reveal rich physics that warrant detailed investigation. In contrast to their parametric study, based solely on numerical simulations, here we have rigorously revealed the instability mechanism using a linear stability analysis and characterized the resulting planar swimming modes using numerical simulations and scaling analysis. 

We have only accounted for the local contributions of SBT and SPT in this study. Accounting for the nonlocal contribution in SBT does not significantly affect our findings, but it does in SPT. Future works will focus on extending our framework to account for the nonlocal contributions as well as inter-filament chemical and hydrodynamic interactions, which are essential for investigating the dynamics of highly flexible active filaments and their collective behavior. The addition of thermal fluctuations in our simulations could also significantly modify the dynamics of active filaments by continuously changing their conformations over time. Semiflexible chains made of isotropic autocatalytic monomers have been observed to exhibit spontaneous deformations and an enhanced diffusivity due to the active motion of the filaments \citep{biswasLinkingIsotropicColloids2017}.

Achieving efficient self-propulsion at microscopic scales, particularly in complex and confined environments, remains a key interdisciplinary challenge. The dynamics of flexible autophoretic filaments explored here present a promising step toward addressing this challenge. The adaptivity provided by the flexible nature of filaments enables motion through confinement -- a feature that living organisms like bacteria and spermatozoa rely on for navigating tight spaces. Furthermore, as in some organisms, flexibility offers additional degrees of freedom, allowing switching between different modes and empowering the selection of functions such as pumping, navigation, or mixing. Our findings thus provide physical insight that can aid future experiments and the design of reconfigurable synthetic active systems having such multifunctional capabilities.

\section*{Acknowledgments}
The authors thank Ondrej Maxian and Sébastien Michelin for feedback on this work. The authors also express their gratitude to Matthew Butler, Benjamin Walker, and Tom Montenegro-Johnson for discussions on the topic. U.M. is grateful to Ondrej Maxian and Brennan Sprinkle for help with some aspects of the numerical approach. A.V. acknowledges funding from the Alexander von Humboldt-Stiftung. P.K. acknowledges funding from the Research and Innovation Foundation of the Republic of Cyprus
under Contract No. EXCELLENCE/0524/0363 for the Excellence Hub project “MICROFIBRES".  

\section*{Data availability}
All the data that support the findings of this study are available from the corresponding author upon reasonable request.

\section*{Code availability}
All the codes that support the findings of this study are available from the corresponding author upon reasonable request.  

\section*{Declaration of interests}
The authors report no conflict of interest.

\appendix

\section{Convergence of the numerical scheme}
\label{appendix: convergence_study}
We conduct a convergence study to verify the accuracy of our numerical scheme. Our numerical results were validated against theoretical predictions in Section \ref{section:validation_of_the_numerical_approach}. As a test case, we consider the buckling instability of an active elastic filament for $\alpha=150$. All simulations are performed for $\rho(s)=\sqrt{1-s^2}$, $\mathcal{AM}=-1$, and $\varepsilon=10^{-2}$.

\begin{figure}[ht]
    \centering
    \includegraphics[width=0.95\linewidth]{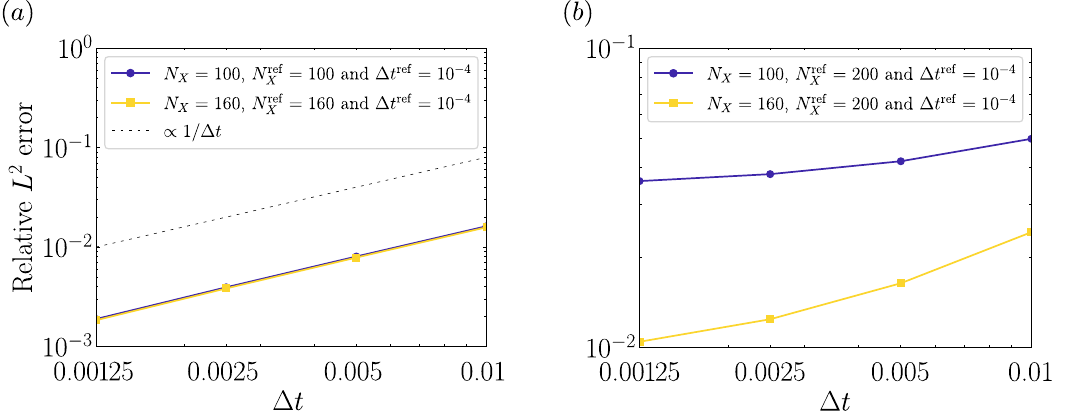}
    \caption{Convergence study. $(a)$ Temporal convergence for different spatial resolutions $N_X$. The reference solution is computed with $N_X^\mathrm{ref}=N_X$ and $\Delta t^\mathrm{ref} = 10^{-4}$. $(b)$ Spatiotemporal convergence. The reference solution is computed with $N_X^\mathrm{ref}=200$ and  $\Delta t^\mathrm{ref} = 10^{-4}$.}
    \label{fig: convergenceStudy}
\end{figure}

To study temporal convergence, we perform -- for $N_X=100$ and $N_X=160$ -- long-time simulations (sufficiently long to reach the nonlinear regime) spanning $t=0$ to $t=15$, using successively increasing time steps: $\Delta t = 0.00125,\; 0.0025,\; 0.005,\; \text{and}\; 0.01$. We measure the maximum relative $L^2$ errors in the centerline positions, $\bm{X}$, against a more refined solution -- computed with $N_X^\mathrm{ref}=N_X$ and $\Delta t^\mathrm{ref}=10^{-4}$. We obtain a first-order temporal convergence, as shown in Fig.\ref{fig: convergenceStudy}$(a)$, thereby validating our temporal discretization.

Thus, we conduct a spatiotemporal convergence, where $L^2$ errors are measured against a reference solution -- computed with $N_X^\mathrm{ref}=200$ and $\Delta t^\mathrm{ref}=10^{-4}$. Our numerical scheme converges with three digits of accuracy, see Fig.\ref{fig: convergenceStudy}$(b)$. The spatiotemporal accuracy is bottlenecked by spatial accuracy. Indeed, for $\rho(s)=\sqrt{1-s^2}$ and a constant $\mathcal{AM}$, the tangential component of the phoretic slip velocity, $\tilde{u}$, exhibits a singularity at the endpoints -- see Eq.\eqref{eq:slip_spheroid_tangent}. Consequently, we obtain a second-order algebraic convergence rather than a spectral convergence in space. In this paper, we used a Chebyshev basis of the first kind, i.e., that excludes the endpoints ($s=\pm1$). Although this does not fully resolve the issue -- $\tilde{u}$ remains nearly singular in the vicinity of the endpoints --, we obtain better accuracy in our simulations with $N_X=101$ and $\Delta t=10^{-4}$.

\bibliography{draft_PRF}

\end{document}